\documentclass[aps,prc,twocolumn,showpacs,preprintnumbers,amsmath,amssymb,nofootinbib]{revtex4-1}

\usepackage{sublabel}
\usepackage[]{graphicx}
\usepackage{epsf}
\usepackage{wrapfig}
\usepackage{epsfig}
\usepackage{wasysym}
\usepackage{color}
\usepackage{pstricks}
\usepackage{amssymb, graphics}
\usepackage{indentfirst}
\usepackage{fancyhdr}
\usepackage{slashed}
\usepackage[latin1]{inputenc}
\usepackage[english]{babel}
\usepackage{amssymb}  
\usepackage{amsmath}  
\usepackage{latexsym} 
\usepackage{amsthm}   

\usepackage{color}

\begin{document}
\title{Density-dependent nucleon-nucleon interaction from three-nucleon forces}
\author{Alessandro Lovato$^{1}$}
\author{Omar Benhar$^{2,3}$}
\author{Stefano Fantoni$^{1,4}$}
\author{Alexey Yu. Illarionov$^{5}$}
\author{Kevin E. Schmidt$^{6}$}

\affiliation
{
$^1$ SISSA and INFN, Sezione di Trieste. I-34014 Trieste, Italy \\
$^2$ INFN, Sezione di Roma. I-00185 Roma, Italy\\
$^3$ Dipartimento di Fisica, Universit\`a ``La Sapienza''. I-00185 Roma, Italy \\
$^4$ CNR-DEMOCRITOS National Supercomputing Center. I-34014 Trieste, Italy\\
$^5$ Dipartimento di Fisica, Universit\`a di Trento. I-38123 Povo, Trento, Italy \\
$^6$ Department of Physics, Arizona State University, Tempe, AZ 85287 \\
}
\date{\today}
\begin{abstract}

Microscopic calculations based on realistic  nuclear hamiltonians, while yielding accurate results for the energies
of the ground and low-lying excited states of nuclei with $A \leq 12$, fail to reproduce the empirical equilibrium properties of 
nuclear matter, that are known to be significantly affected by three-nucleon forces. 
We discuss a scheme suitable to construct a density-dependent two-nucleon potential, in which the effects of 
$n$-particle interactions can be included by integrating out the degrees of freedom  of $(n-2)$-nucleons.
Our approach, based on the formalism of correlated basis function and state-of-the-art
models of the two- and three-nucleon potentials, leads to an effective interaction 
that can be easily employed in nuclear matter calculations, yielding results in good agreement with those
obtained from the underlying three-body potential.
\end{abstract}


\maketitle


\section{Introduction}

The results of ab initio microscopic calculations consistently suggest that realistic  nuclear hamiltonians, 
including both two- and three-nucleon potentials, while providing a quantitative account of the energies of the 
ground and low-lying excited  states of nuclei with $A\leq12$ \cite{steve_bob, steve_RNC}, fail to explain the 
empirical equilibrium properties of nuclear matter. 
This problem can be largely ascribed to the uncertainties associated with the description of 
three-nucleon interactions, whose contribution turns out to be significant. 

A signal of the limitations of the commonly employed three-nucleon potential models (e.g. the Urbana IX model of Ref. \cite{UIX})
has been recently provided by the authors of Ref. \cite{AFDMC1}, who carried out a study of symmetric nuclear matter within the Auxiliary Field Diffusion Monte Carlo (AFDMC) approach. Their results, obtained using a truncated version \cite{V8P}
of the state-of-the-art  nucleon-nucleon potential of Ref. \cite{av18}, show that AFDMC simulations do not lead to an increase of the binding 
energy predicted by Fermi-Hyper-Netted-Chain 
(FHNC) and Brueckner-Hartree-Fock (BHF) calculations \cite{FHNC-BHF}. 

Different three-nucleon potential models \cite{UIX,ILL}, yielding similar results when applied to the calculation of nuclear properties, predict sizably different equations of state (EoS) of pure neutron matter at zero temperature and densities 
exceeding the nuclear matter saturation density, $\rho_0 = 0.16$ fm$^{-3}$ \cite{AFDMC2}. In this region, the three-nucleon 
force contribution to the binding energy becomes very large, the ratio between the potential energies associated 
with two- and the three-body interactions being $\sim 20 \%$ at density $\rho \sim 2 \rho_0$ (see, e.g. Ref. \cite{akmal}). 
The size of the three-body force contribution suggests that, at large $\rho$, interactions involving four or more nucleons may also 
play an important role, and should be taken into account.

In view of the severe difficulties involved in the implementation of the existing models of three-nucleon interactions  
in many-body calculations, the explicit inclusion of  four- and more-body potentials does not appear to be a viable 
option. In this paper we follow a different strategy, somewhat along the line of the Three-Nucleon-Interaction (TNI) model 
proposed by Lagaris and Pandharipande \cite{LP1} and Friedman and Pandharipande \cite{FrP} in the 1980s.  

The authors of Refs. \cite{LP1,FrP} suggested that the main effects of three- and many-nucleon forces  can be taken 
into account through an effective, density-dependent two-nucleon potential. However, they adopted a purely 
phenomenological procedure, lacking a clearcut interpretation based on the the analysis of many-nucleon interactions at microscopic level. 

The TNI potential consists of two density-dependent functions involving three free parameters, whose values were determined through a fit of the saturation density, binding energy per nucleon and compressibility of symmetric nuclear matter (SNM), obtained from FHNC variational calculations. The numerical values of the three model parameters resulting from recent calculations performed by using AFDMC simulations turn out to be only marginally different from those of the original TNI potential \cite{Gand2010}.

The TNI potential has been successfully applied to obtain a variety of nuclear matter properties, such as  the nucleon momentum distribution \cite{mom_dis}, the linear response \cite{response_1,response_2}, and the Green's function \cite{spec_func1, spec_func2}, within the Correlated Basis Function (CBF) approach (for a review of CBF theory and its applications, see Ref.\cite{fabrocini_fantoni} and references therein). 

The strategy based on the development of two--body density-dependent potentials has been later abandoned, because their application to the study of finite nuclei involves a great deal of complication, mainly stemming from the breakdown of translation invariance. While in uniform matter the density is constant and the expansion of the effective potential in powers of $\rho$ is straightforward, in nuclei different powers of the density correspond to different operators, whose treatment is highly non trivial.

However, the recent developments in numerical methods for light nuclei seem to indicate that the above difficulties may turn out to be much less severe then those implied in the modeling of explicit many--body forces and, even more, in their use in {\sl ab initio} nuclear calculations.

In view of the observation, based on a variety of experimental evidence \cite{RMP1,Pand1997}, that short range nucleon--nucleon (NN) correlations are a 
fundamental feature of nuclear structure, the description of nuclear dynamics in terms of interactions derived in coordinate space, like the Urbana-Argonne models, appears to be the most appropriate, for both conceptual and technical reasons.

First of all, correlations between nucleons are predominantly of spatial nature, in analogy with what happens in all known strongly correlated systems. 
In addition, one needs to clearly distinguish the effects due to the short--range repulsion from those due to relativity. Finally, quantum Monte Carlo methods have serious difficulties in dealing with highly non local interactions. For all the above reasons we stick to two--body density-dependent potentials of the Urbana-Argonne type.

Our approach is based on the tenet that $n$-body potentials ($n\geq3$) can be replaced by an effective  two-nucleon potential, obtained  through  an average over the degrees of freedom of $n-2$ particles. Hence, the effective potential can be written as a sum of contributions ordered according to powers of density, the $p$-th order term being associated with $(p+2)$-nucleon forces.

Obviously, such an approach requires that the average be carried out using a formalism suitable to account for the  
full complexity of nuclear dynamics. Our results show that, in doing such reduction, of great importance is the proper inclusion 
of both dynamical and statistical NN correlations, whose effects on many nuclear observables have been found to be large  \cite{RMP1,Pand1997}. 

In this work, we use CBF and the Fantoni-Rosati (FR) cluster expansion formalism \cite{FR_cluster,fabrocini_fantoni,ariasdesaavedra_bisconti_co_fabrocini} 
to perform the calculation of the terms linear in density of the effective potential, arising from the irreducible 
three-nucleon interactions modeled by the UIX potential.

It should be noticed that our approach significantly improves on the TNI model, as the resulting potential is obtained from a realistic microscopic three-nucleon force, which provides an accurate description of the properties of light nuclei.

While being the first step on a long road, the results discussed in this paper are valuable in their own right, as the effective potential can be easily implemented in the AFDMC computational scheme to obtain the EoS of SNM. Similar calculations using the UIX potential are not yet possible, due to the complexities arising from the commutator 
term. In addition, the density-dependent potential can be used to include the effects of three-nucleon interactions in the 
calculation of the nucleon-nucleon scattering cross section in the nuclear medium. The knowledge of this 
quantity is required to obtain a number of nuclear matter properties of astrophysical interest, ranging from the 
transport coefficients to the neutrino emission rates \cite{BV,BFFV}.

In Section~\ref{TBF} we discuss the main features of the existing theoretical models of the three-nucleon force, 
while Section~\ref{many-body} is devoted to a brief review of the many-body approach based on CBF and the cluster 
expansion technique.  In Section~\ref{ddpot} we describe the derivation of the density-dependent interaction, pointing out the role of dynamical and statistical correlation effects. In Section~\ref{Num_Calc} we compare the energy per particle of nuclear matter obtained from the effective potential to that resulting from highly refined calculations, carried out using the Argonne $v_{6}^\prime$ \cite{V6P} and $v_{8}^\prime$ \cite{V8P} nucleon-nucleon potentials and the Urbana IX three-nucleon potential \cite{UIX}. Finally, in Section~\ref{conclusions} we summarize our findings and state the conclusions.


\section{Three nucleon forces}
\label{TBF}

Nuclear many-body theory (NMBT) is based on the assumption that nuclei can be
described in terms of point like nucleons of mass $m$, whose dynamics are dictated by
the hamiltonian
\begin{equation}
\hat{H} = \sum_i -\frac{{\nabla}^2_i}{2m} + \sum_{j>i} \hat{v}_{ij} + \sum_{k>j>i} \hat{V}_{ijk}  \ .
\label{hamiltonian}
\end{equation}

Before describing the three nucleon potential $V_{ijk}$, let us discuss the $v_{8}$ two-body potential model, that will be used throughout the paper. It is given by 
\begin{equation}
\hat{v}_{ij}=\sum_{p=1}^8 v^p(r_{ij})O^{p}_{ij}\, ,
\label{eq:av8_potential}
\end{equation}
where
\begin{equation}
O^{p=1-8}_{ij}=(1,\sigma_{ij},S_{ij},\mathbf{L}_{ij}\cdot\mathbf{S}_{ij})\otimes(1,\tau_{ij})\,.
\end{equation}
In the above equation,  $\sigma_{ij}={\boldsymbol \sigma}_i \cdot {\boldsymbol \sigma}_j$ and 
$\tau_{ij}={\boldsymbol \tau}_i \cdot {\boldsymbol \tau}_j$, where ${\boldsymbol \sigma}_i$ and ${\boldsymbol \tau}_i$ are Pauli matrices acting on the
spin or isospin of the $i$-th, while
\begin{equation}
S_{ij}=T_{ij}^{\alpha\beta}\sigma_{i}^\alpha\sigma_{j}^\beta=(3\hat{r}_{ij}^\alpha\hat{r}_{ij}^\beta-\delta^{\alpha\beta})\sigma_{i}^\alpha\sigma_{j}^\beta \ ,
\end{equation}
with $\alpha, \ \beta= 1, \ 2, \ 3$, is the tensor operator, $\mathbf{L}_{ij}$ is the relative angular momentum 
\begin{equation}
\mathbf{L}_{ij}=\frac{1}{2i}(\mathbf{r}_i-\mathbf{r}_j)\times (\boldsymbol{\nabla}_i-\boldsymbol{\nabla}_j)
\end{equation}
and $\mathbf{S}_{ij}$ is the total spin of the pair
\begin{equation}
\mathbf{S}_{ij}=\frac{1}{2}(\boldsymbol{\sigma}_i+\boldsymbol{\sigma}_j)\, .
\end{equation}

Such potentials have exactly the same form as the first eight components of the state-of-the-art Argonne $v_{18}$ potential \cite{av18}. We will be using the so called Argonne $v_{8}^\prime$ and Argonne $v_{6}^\prime$ potentials, which are not simple truncations of the Argonne $v_{18}$ potential, but rather reprojections \cite{V6P}. 

The Argonne $v_{8}^\prime$ potential is obtained by refitting the scattering data in such a way that all $S$ and $P$ partial waves as well as the $^3 D_1$ wave and its coupling to $^3 S_1$ are reproduced equally well as in Argonne $v_{18}$. In all light nuclei and nuclear matter calculations the results obtained with the $v_{8}^\prime$ are very close to those obtained with the full $v_{18}$, and the difference $v_{18}-v_{8}^\prime$ can be safely treated perturbatively.

The Argonne $v_{6}^\prime$ is not just a truncation of $v_{8}^\prime$, as the radial functions associated with the first six operators are adjusted to preserve the deuteron binding energy. Our interest in this potential is mostly due to the fact that AFDMC simulations of nuclei and nuclear matter can be performed most accurately with $v_6$--type of two--body interactions. Work to include the spin--orbit terms in AFDMC calculations is in progress. On the other hand we need to check the accuracy of our proposed density-dependent reduction with both FHNC and AFDMC many--body methods before proceeding to the construction of a realistic two--body density-dependent model potential and comparing with experimental data.

It is well known that using a nuclear Hamiltonian including only two-nucleon interactions leads to the underbinding of light nuclei and overestimating the equilibrium density of nuclear matter. Hence, the contribution of three-nucleon interactions must necessarily be taken into account, 
by adding to the Hamiltonian  the corresponding potential,  e.g. the widely used Urbana IX (UIX) \cite{UIX}. 

The potential of Ref. \cite{UIX} consists of two terms. The attractive two-pion ($2\pi$) exchange  interaction $V^{2\pi}$ turns out to be helpful in fixing the problem of light nuclei, but makes the nuclear matter energy worse. The purely phenomenological repulsive term $V^R$ prevents nuclear matter from being overbound at large density.

\begin{figure}[!h]
\vspace{0.2cm}
\begin{center}
\includegraphics[angle=0,width=7.0cm]{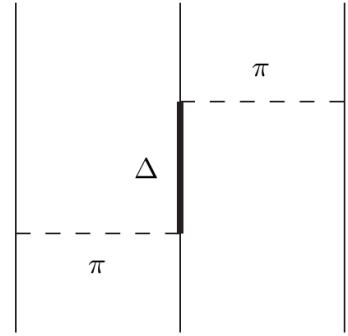}
\caption{Feynman Diagram associated with the Fujita Miyazawa three-nucleon potential term.}
\label{fig:Fujita_Miyazawa}
\end{center}
\end{figure}

The $V^{2\pi}$ term was first introduced by Fujita and Miyazawa \cite{Fujita_Miyazawa} to describe the process 
whereby two pions are exchanged among nucleons and a $\Delta$ resonance is excited in the intermediate state, as 
shown in the Feynman diagram of Fig.~\ref{fig:Fujita_Miyazawa}. It can be conveniently written in the form
\begin{equation}
\hat{V}^{2\pi}=\sum_{cyclic}\hat{V}^{2\pi}_{ijk}=A^{2\pi}\sum_{cyclic}\hat{O}_{ijk}^{2\pi}\, , 
\end{equation}
where
\begin{align}
\hat{O}_{123}^{2\pi}&=A_{2\pi}\Big(\{\hat{X}_{12},\hat{X}_{23}\}\{\tau_{12},\tau_{23}\}\nonumber\\
&+\frac{1}{4}[\hat{X}_{12},\hat{X}_{23}][\tau_{12},\tau_{23}]\Big)
\label{eq:op_structure_UIX}
\end{align}
and
\begin{equation}
\hat{X}_{ij}=Y(m_\pi r)\sigma_{ij}+T(m_\pi r)S_{ij}\, .
\end{equation}
The radial functions associated with the spin and tensor components read
\begin{align}
Y(x)&=\frac{e^{-x}}{x}\xi_Y(x)\\
T(x)&=\Big(1+\frac{3}{x}+\frac{3}{x^2}\Big)Y(x)\xi_T(x)\, ,
\label{eq:YT}
\end{align}
while the $\xi(x)$ are short-range cutoff functions defined by
\begin{equation}
\xi_{Y}(x)=\xi_{T}(x)=1-e^{-cx^2}\, .
\end{equation}
In the UIX model, the cutoff parameter 
is kept fixed at $c=2.1$ fm$^{-2}$, the same value as in the
cutoff functions appearing in the one-pion exchange term of the Argonne $v_{18}$ two-body potential. On the other hand, $A_{2\pi}$ is varied to fit the observed binding energies of $^3$H and $^4$He. The three-nucleon interaction depends on the choice of the NN potential; for example, using the Argonne $v_{18}$ model one gets $A_{2\pi}=-0.0293\,\,\text{MeV}$.

The repulsive term $V^R$ is spin-isospin independent and can be written in the simple form
\begin{equation}
V^R=\sum_{cyclic}V^{R}_{ijk}=U_0\sum_{cyclic}T^2(m_\pi r_{ij})T^2(m_\pi r_{jk})\, ,
\label{eq:VR}
\end{equation}
with $T(x)$ defined in Eq. (\ref{eq:YT}). The strength $U_0$,  adjusted to reproduce the empirical nuclear matter 
saturation density, is $U_0=0.0048\,\,\text{MeV}$ with $v_{18}$.

The two parameters $A_{2\pi}$ and $U_0$ have different values  for $v_{8}^\prime$ and $v_{6}^\prime$. We disregard such differences in the present analysis, mostly aimed at testing the quality of our density-dependent reduction of the UIX three--body potential, rather than reproducing empirical data. 
\\

\section{formalism}
\label{many-body}

\subsection{Correlated basis theories}

One of the most prominent features of the nucleon-nucleon (NN) interaction is the presence of a repulsive core, giving rise to 
strong correlations that cannot be taken into account within the independent particle picture. 

This problem has long been recognized, and was clearly pointed out by Blatt and Weisskopf over fifty years ago. In their 
classic Nuclear Physics book, first published in 1952, they warn the reader that ``the limitation of any independent particle 
model lies in its inability to encompass the correlation between the positions and spins of the various particles in the 
system'' \cite{BW}.

Let us consider uniform nuclear matter, defined as a translationally invariant  system of protons and neutrons, 
in which the electromagnetic interaction is turned off. In the absence of interactions, such a system can be described as a 
Fermi gas at zero temperature, and its ground state wave function reduces to the antisymmetrized product (Slater determinant) of 
 orbitals associated with the single particle states belonging to the Fermi sea: 
\begin{equation}
\Phi(x_1,\dots,x_A)=\mathcal{A}[\,\phi_{n_1}(x_1)\dots\phi_{n_A}(x_A)\,]\, ,
\label{def:Phi}
\end{equation}
with
\begin{equation}
\phi_{n_i}(x_i)\equiv\phi_{\mathbf{k_i},\sigma_i \tau_i}(\mathbf{r}_i)= \varphi_{{\bf k}_i} ({\bf r}_i) \chi_{\sigma_i} 
\eta_{\tau_i} \, ,
\end{equation}
and 
\begin{equation}
\varphi_{{\bf k}_i} ({\bf r}_i)= \frac{1}{\sqrt{\Omega}} {\rm e}^{i\mathbf{k_i}\mathbf{r}_i} \ .
\end{equation}
In the above equations, $\Omega$ is the normalization volume, $\chi_{\sigma_i}$ and $\eta_{\tau_i}$ are Pauli spinors, describing the nucleon spin and isospin and $|\mathbf{k}_i|<k_F=(6\pi^2\rho/\nu)^{1/3}$. Here $k_F$ is the Fermi momentum, while $\rho$ and $\nu$ denote the density and the degeneracy of the momentum eigenstates [$\nu=$2, 4 pure neutron matter (PNM) and symmetric nuclear matter (SNM), respectively]. The generalized coordinate $x_i\equiv\{\mathbf{r}_i,\sigma_i,\tau_i\}$ represents both the position and the spin-isospin variables of the $i$-th nucleon, while $n_i$ denotes the set of quantum numbers specifying the single particle state. 

The antisymmetrization operator $\mathcal{A}$ can be written in the form 
\begin{equation}
\mathcal{A}=1-\sum_{i<j} P_{ij}+\sum_{i<j<k}(P_{ij}P_{jk}+P_{ik}P_{kj})+\dots\,  \  ,
\label{eq:anti_exch}
\end{equation}
where
\begin{equation}
P_{ij}=\frac{1}{4}(1+\sigma_{ij})(1+\tau_{ij})\exp[-i(\mathbf{k}_i-\mathbf{k}_j)\cdot \mathbf{r}_{ij}]\,  
\label{def:Pij}
\end{equation}
is the two-particle exchange operator, defined by the relation
\begin{equation}
P_{ij} \phi_{n_i}(x_i) \phi_{n_j}(x_j) = \phi_{n_i}(x_j) \phi_{n_j}(x_i) \ .
\end{equation}
Note that, as shown by Eq.(\ref{def:Pij}), the exchange operators act on both the radial and spin-isospin components 
of the nucleon wave function.

Due to the strong repulsive core, the matrix elements of $v_{ij}$ between eigenstates of the non interacting system
\begin{equation}
\langle\phi_{\mathbf{k}_{1'}\sigma_{1'}\tau_{1'}}\phi_{\mathbf{k}_{2'}\sigma_{2'}\tau_{2'}}|v_{12}|\phi_{\mathbf{k}_{1}\sigma_{1}\tau_{1}}\phi_{\mathbf{k}_{2}\sigma_{2}\tau_{2}}\rangle
\end{equation}
turn out to be very large, or even divergent if the core of the NN potential is infinite. As a consequence, perturbative calculations carried out using the {\em bare} NN potential and the Fermi gas basis states are unavoidably plagued by lack of convergence.

To circumvent this problem, one can follow two different strategies, leading to either G-matrix or CBF perturbation theory. Within the former approach, the bare potential $v_{ij}$ is replaced by  a well behaved effective interactions, the so called G-matrix, which is obtained by summing up the series of particle--particle ladder diagrams. In the second approach, nonperturbative effects are handled through a change of basis functions.  
   
Correlated basis theories of Fermi liquids\cite{fabrocini_fantoni,ariasdesaavedra_bisconti_co_fabrocini,clark,OCBfantoni} are a natural extension of variational approaches in which the trial ground state wave function is written in the form 
\begin{equation}
|\Psi_0\rangle=\frac{\hat{F}|\Phi_0\rangle}{\langle\Phi_0|\hat{F}^\dagger \hat{F}|\Phi_0\rangle^{1/2}}\, .
\end{equation}
In the above equation, $\hat{F}$ is a correlation operator, whose structure reflects the complexity of the nucleon-nucleon  potential \cite{wiringa_pandha_1}:
\begin{equation}
F=\mathcal{S} \prod_{j>i=1}^A F_{ij} \ ,
\label{eq:Foperator}
\end{equation}
with
\begin{equation}
\hat{F}_{ij} = \sum_{p=1}^6 f^{p}(r_{ij})\hat{O}^{p}_{ij} \ .
\end{equation}
Note that the symmetrization operator $\mathcal{S}$ is needed to fulfill the requirement of antisymmetrization of the state 
$|\Psi_n\rangle$, since, in general, $[\hat{O}^{p}_{ij},\hat{O}^{q}_{ik}] \neq 0$.
The correlated basis (CB) is defined as
\begin{equation}
|\Psi_n\rangle=\frac{\hat{F}|\Phi_n\rangle}{\langle\Phi_n|\hat{F}^\dagger \hat{F}|\Phi_n\rangle^{1/2}}\, ,
\end{equation}
where $|\Phi_n\rangle$ is a generic {\em n} particle -- {\em n} hole Fermi gas state. The CB states are normalized but not orthogonal to each other. They have been used within non orthogonal perturbation theory \cite{clark,krotscheck} to study various properties of quantum liquids. An exhaustive analysis of the convergence properties of this perturbation scheme has never been carried out, but the truncation of the series at a given perturbative order is known to lead to nonorthogonality spuriosities, whose effects are not always negligible. A much safer and efficient procedure,  in which one first orthogonalizes the CB states by using a combination of Schmidt and L$\ddot{o}$wdin transformations and then uses standard perturbation theory, has been proposed by Fantoni and Pandharipande \cite{OCBfantoni}.

The radial functions $f^{p}(r_{ij})$, appearing in the definition of the correlation operator are determined by the minimization of the  energy expectation value
\begin{equation} 
 E_V=\langle\Psi_0|H|\Psi_0\rangle \, , 
\label{eq:ev}
\end{equation}
which provides an upper bound to the true ground state energy $E_0$. In principle, that can be done by solving the Euler equations resulting from 
the functional minimization of $E_V$ with respect to the correlation functions $f^p(r_{ij})$, in analogy with what has been done in Jastrow theory 
of liquid $^3$He. However, the presence of the spin-isospin dependent correlation operators and their non--commutativity makes the application of 
this procedure to nucleonic systems almost prohibitive. In this case the functional minimization can be carried in a more straightforward fashion on the 
lowest order cluster contribution to  $E_V$, paying the price of introducing proper constraints and the associated variational parameters, as discussed 
below.
\subsection{Cluster expansion}
\label{cluster:expansion}

In CBF theories the calculation of $E_V$ is carried out by i) expanding the r.h.s. of eq. (\ref{eq:ev}) in powers of proper expanding functions that vanish in uncorrelated matter and ii) summing up the main series of the resulting cluster terms by solving a set of coupled integral equations. The FR cluster expansion \cite{FR_cluster} has been derived to accomplish the first of these two steps for the case of Jastrow correlated models. It has been obtained through a generalization 
of the concepts underlying the Mayer expansion scheme,  originally developed to  describe classical liquids \cite{mayer}, to the case of quantum Bose and Fermi systems. In this case the expanding quantity is given by
\begin{equation}
h(r_{ij})={f^{c}(r_{ij})}^2-1\, ,
\label{eq:hcentral}
\end{equation}
where $f^c$ is the only correlation of the Jastrow model. Notice that in the calculation of $E_V$ the kinetic energy operators ${\nabla_i}^2$ also act on the correlation functions, giving rise to additional expanding quantities. For the sake of simplicity, and since here we are interested in calculating the expectation 
values of two- and three-body potentials, we will not address this issue. It has been proved \cite{FR_cluster} that the FR cluster expansion is linked, and 
therefore does not suffer the appearance of infinities in the thermodynamic limit.
The FR techniques have been subsequently extended and extensively used  to deal with spin--isospin dependent correlation operators, like those of 
Eq. (\ref{eq:Foperator}) \cite{wiringa_pandha_1,wiringa_pandha_2}. In this case, besides the expanding function $h(r_{ij})$ of eq. (\ref{eq:hcentral}) one 
has to also consider the following ones:
\begin{equation}
2f^{c}(r_{ij})f^{p>1}(r_{ij}) \quad,\quad f^{p>1}(r_{ij})f^{q>1}(r_{ij})\, .
\label{eq:hoperatorial}
\end{equation} 

\begin{figure}[!ht]
\vspace{0.2cm}
\begin{center}
\includegraphics[angle=0,width=6.0cm]{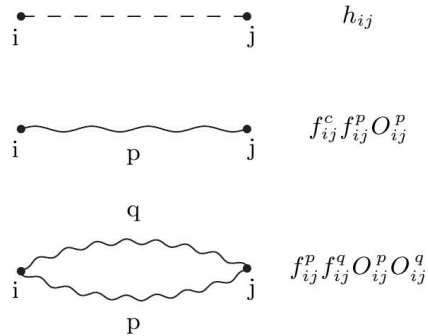}
\caption{Different kinds of correlation bonds.}
\label{fig:pass_corr_lines}
\end{center}
\end{figure}

The cluster terms are most conveniently represented by diagrams \cite{wiringa_pandha_1,fabrocini_fantoni}. The diagrammatic representation of the above expanding functions is given by the bonds displayed in Fig. \ref{fig:pass_corr_lines}, in which $h_{ij}$ is represented by a dashed line, 
$2f^{c}_{ij}f^{p}_{ij}$ by a single wavy line, and $f^{p>1}_{ij}f^{q>1}_{ij}$ by a doubly wavy line .

The $h(r_{ij})$ is the largest of all the expanding quantities and the cluster terms (and similarly the corresponding cluster diagrams)
involving  these functions (or bonds) have to be summed up as massively as possible. This can be accomplished by solving the FHNC equations of 
Ref. \cite{fantoni_rosati}, which already in their basic form sum up all diagrams at all orders,  with the exception of the so called elementary diagrams. 

The cluster diagrams involving operatorial bonds, like those representing the functions given in Eq. (\ref{eq:hoperatorial}), cannot be summed up as massively as the {\sl scalar} diagrams of FHNC type. This is due to the additional complexity associated with the non commutativity of the spin--isospin 
dependent correlation operators. The most powerful summation scheme which has been derived so far is the so called 
Single Operator Chain of Refs. (\cite{wiringa_pandha_1,wiringa_pandha_2}), generally denoted as FHNC/SOC approximation. 

For the sake of clarity,  in the following we summarize the main features of the FR cluster expansion and the FHNC/SOC approximation, extensively reviewed in  \cite{wiringa_pandha_1,fabrocini_fantoni}. 

Let us consider the expectation value of the NN potential. Exploiting the symmetry properties of the wave function it can be written in the form
 \begin{eqnarray}
\langle \hat{v} \rangle &=& \frac{A(A-1)}{2}\frac{\langle \Phi_0^* | \hat{F}^\dagger \hat{v}_{12} \hat{F} | \Phi_0 \rangle}
{\langle  \Phi_0^* | \hat{F}^\dagger \hat{F} |  \Phi_0\rangle }\, .
\label{eq:two_body_exp}
\end{eqnarray}
Numerator and denominator of the above equation are expanded in powers of the functions defined above. The expansion of $\hat{F}^\dagger \mathcal{O}\hat{F}$, with $\mathcal{O}=\hat{v}_{12}, \openone$ for the numerator and the denominator, leads to series of terms, 
say $\hat{X}_n^{(N,D)}$, where the labels {\textit N} and {\textit D} stand for numerator and denominator, respectively, each characterized by the 
number $n$ of correlated nucleons, i.e. those appearing in the argument of the expanding functions. Integrating such terms over the variables of the remaining uncorrelated nucleons amounts to multiplying $\hat{X}_n$ by the $n$--body Fermi distribution operator $\hat{g}_F(1,\ldots ,n)$. 
Consider for instance one of the cluster terms of the numerator, whose structure is given by
\begin{align}
\langle \hat{X}_n\rangle &= {A\choose{n}}\frac{\int dX \ \Phi_0^*\hat{X}(1,\ldots,n) \Phi_0}{\int dX \ \Phi_0^*\Phi_0}\nonumber \\
&= \frac{\rho^n}{n!}\int dx_1\ldots dx_n \hat{X}(1,\ldots,n) \hat{g}_F(1,\ldots,n)\, .
\label{eq:cluster_term}
\end{align}
where $dX\equiv dx_1 \ldots dx_A$ and $\int dx_i$ stands for integration over the coordinate $\vec{r}_i$ and tracing over the spin and isospin variables of the $i$--th nucleon, and 
\begin{eqnarray}
\hat{g}_F(1,\ldots,n) &=& \frac{A\ldots (A-n+1)}{\rho^n} \nonumber \\
&\times& \frac{\int dx_{n+1}\ldots dx_A\Phi_0^*\Phi_0}{\int dX \ \Phi_0^*\Phi_0}\, .
\label{eq:FDO}
\end{eqnarray}
Note that $\hat{X}(1,\ldots,n)$ can be moved to the left of $\Phi_0^*$ to 
obtain the second line of Eq. (\ref{eq:cluster_term}) as, on account of the property
 \begin{align}
&{A\choose{n}}\int dX\, \Phi_0^*\hat{X}(1,\ldots,n) \Phi_0 \nonumber \\
&= \sum_{n_1,\dots,n_n}\int d\vec{r}_{1}\ldots d\vec{r}_n [\phi_{n_1}^*(x_1)\dots \phi_{n_n}^*(x_n)]\nonumber \\
&\frac{\hat{X}(1,\ldots,n)}{n!} \mathcal{A}[\phi_{n_1}(x_1)\dots \phi_{n_n}(x_n)]\ , 
\label{eq:def_antisimm}
\end{align}
one needs to  antisymmetrize  $\Phi_0$ only.

Summing over the states belonging to the Fermi sea for each $n_i$ independently leads to an expression which does not depend on the number of particle $A$. There is no violation of the Pauli principle because of the antisymmetrization of $\Phi_0$. More specifically, each term of the r.h.s. of Eq. (\ref{eq:def_antisimm}) coming from the antisymmetrization of $\Phi_0$ is Pauli violating, but the total sum it is not. On the other side,  the independence of $A$ has very useful consequences. One of them is that the numerator of $\langle v \rangle$  can be easily recognized as the product of the denominator times the sum of linked cluster terms. In addition, the FR cluster expansion is exact for any number of particles, not just in the thermodynamic limit like, for example, the Mayer expansion. This property has been exploited in FHNC calculations of finite nuclear systems like nuclei \cite{ariasdesaavedra_bisconti_co_fabrocini} or nucleon 
confined in periodical box \cite{PBFHNC}.

The operatorial $n$--body Fermi distribution function  $\hat{g}_F(1,\ldots,n)$ includes a direct term corresponding to $1$ in Eq. (\ref{eq:anti_exch}) and a number of exchange terms generated according to the algebra of the exchange operators $P_{ij}$. 

The basic statistical (exchange) correlation is described by the Fermi gas one--body density matrix
\begin{eqnarray}
\ell_{ij} &=& \frac{1}{\rho}\sum_n \phi_n^*(\vec{r}_i) \phi_n(\vec{r}_j) \nonumber \\
&=& \frac{1}{\nu} \ell(k_Fr_{ij})\sum_{\chi_\sigma\eta_\tau} \chi_{\sigma_i}\eta_{\tau_i} \chi_{\sigma_j}^\dagger\eta_{\tau_j}^\dagger\, , 
\end{eqnarray} 
where the Slater function $\ell(k_Fr_{ij})$ is given by
\begin{equation}
\ell(k_Fr_{ij}) = 3\Big[\frac{\sin(k_Fr_{ij})-k_Fr_{ij}\cos(k_Fr_{ij})}{(k_Fr_{ij})^3}\Big]\, .
\end{equation}
Diagrammatically, the exchange correlation $\ell_{ij}$, referred to as ``exchange bond'', is represented  by an oriented solid line. The $P_{ij}$--algebra implies that the exchange bonds form closed loops which never touch each other. If $\hat{X}(1,\ldots,n)$ is made of scalar correlations $h(r_{ij})$ only, then all  nucleons in a given exchange loop must be in the same spin--isospin state and the Fermi distribution operators $\hat{g}_F$ of Eq. (\ref{eq:FDO}) reduces to the standard Fermi gas distribution functions. For example, the two--body Fermi distribution function is given by
\begin{equation}
g_F(r_{ij}) = 1-\frac{1}{\nu}l^2(k_Fr_{ij})\, .
\end{equation} 

As an example, consider the two-body cluster contribution. From Eq. (\ref{eq:cluster_term}) it can be written as
\begin{align}
\langle \hat{X}_2\rangle &=\frac{\rho^2}{2}\int dx_1dx_2\hat{X}(1,2)\hat{g}_F(1,2)\nonumber\\
&=\frac{\rho^2}{2}\sum_{n_1,n_2}\int d\vec{r}_1d\vec{r}_2\phi_{n_1}^*(x_1)\phi_{n_2}(x_2)^*\hat{X}(1,2)\nonumber \\
&\times (1-P_{12}) \phi_{n_1}^*(x_1)\phi_{n_2}(x_2)
\end{align}
The sum over the states belonging to Fermi sea implies a sum over the spin-isospin quantum numbers, which amounts to computing the trace 
of the spin and isospin operators appearing in $\hat{X}(1,2)(1-P_{12})$. The trace is normalized to unity, as  summation over the momenta 
leads to the appearance of a factor $(1/\nu)$ in both the direct and exchange term. The final result is
\begin{equation}
\langle \hat{X}_2\rangle =\frac{1}{2}\int dx_1dx_2\hat{X}(1,2)(1-P_{12}\ell^2(k_Fr_{12}))\, .
\end{equation}

\subsubsection{Diagrammatic rules}

The diagrams consist of dots (vertices) connected by different kinds of correlation lines. Open dots represent the active (or interacting) particles ($1$ and $2$), while black dots are associated with passive 
particles, i.e. those in the medium. Integration over the coordinates of the passive particles leads to the appearance of a factor $\rho$.  

\begin{figure}[!ht]
\vspace{0.2cm}
\begin{center}
\includegraphics[angle=0,width=7.0cm]{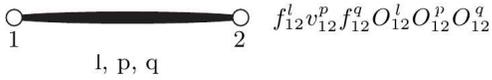}
\caption{Graphical representation of an interaction line.}
\label{fig:act_corr_lines}
\end{center}
\end{figure}

Active correlations must be treated differently from the passive ones, as the components  $v^{\, p}_{12}$ of the two--body potential may be singular, thus leading to divergent integrals. In the diagrammatic expansion of $\langle \hat{v} \rangle/A$, the quantity $F_{12}\hat{v}_{12}F_{12}$  is represented by a thick solid line, denoted as ``interaction line'' and depicted in Fig. \ref{fig:act_corr_lines}.

Exchange lines form closed loops, oriented clockwise or counterclockwise, the simplest of which is the two--body loop yielding  
a contribution $-\ell^2(k_Fr_{ij})/\nu$. In addition to the extra factors coming from the algebra arising from the spin--isospin structure of
 the corresponding cluster term, an $n$--vertex loop carries a factor $(-)(2\nu)(-1/\nu)^n$, where $ -1/\nu$ is associated with each 
 exchange operator $\ell_{ij}$ and $-2\nu$ is due to the presence of $\nu$ spin--isospin species of the loop, combined with the existence of
  $2$ different orientation and to the minus sign coming from the permutations. The two--body loop is an exception to this rule, because there 
  is only one such loop. Therefore, the corresponding factor is $-1/\nu$ rather than $-2/\nu$.

The correlation bonds of Figs. (\ref{fig:pass_corr_lines}\ref{fig:act_corr_lines}) cannot be superimposed to each other. They can only be superimposed to exchange bonds.

The allowed diagrams are all linked, as a result of the linked cluster property discussed above. 

\begin{figure}[!h]
\vspace{0.2cm}
\begin{center}
\includegraphics[angle=0,width=4.5cm]{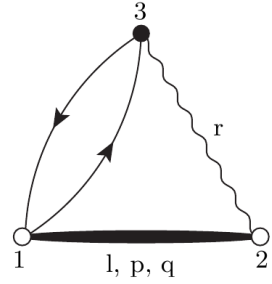}
\caption{Example of diagram appearing in the cluster expansion of $\langle\hat{v}\rangle$.}
\label{fig:ex_cluster_diag}
\end{center}
\end{figure}

A typical diagram of the FR cluster expansion is sketched in Fig. \ref{fig:ex_cluster_diag}. Its contribution to the potential energy expectation value $\langle\hat{v}\rangle$ is given by
\begin{eqnarray}
&&\langle\hat{v}\rangle_{(Fig.\ref{fig:ex_cluster_diag})} = 3 \ \Omega  \ \frac{\rho^3}{3!}\sum_{r,l,p,q}\int d\vec{r}_{12}d\vec{r}_3  \nonumber \\
&&  \left[ -\frac{\ell^2(k_Fr_{13})}{4} \right] f^c(r_{23})f^r(r_{23})f^l(r_{12})v^p(r_{12})f^q(r_{12}) \nonumber \\
&\times&\frac{1}{2} \text{Tr}_{123}\Big[ 4P_{13} \big(\{O^r_{23},O^l_{12}\}O^p_{12}O^q_{12}\nonumber \\
&+& O^l_{12}O^p_{12}\{O^q_{12},O^r_{23}\} \big)\Big]\, ,
\label{eq:ex_cluster_diag}
\end{eqnarray}
where the trace $\text{Tr}_{123}$ is carried out in the spin-isospin spaces of particles 1, 2 and 3. The factor $\Omega$ comes from the relation $\int d\vec{r}_1d\vec{r}_2 =\Omega\int d\vec{r}_{12}$, due to translation invariance, and $3$ is a symmetry factor. 
The four orderings appearing on the r.h.s. of Eq. (\ref{eq:ex_cluster_diag}) correspond to the two possible positions of $f^r(r_{23})$, on either the left-
or right-hand side of the operator $F_{12}\hat{v}_{12}F_{12}$.

\subsubsection{FHNC/SOC approximation}
 
All the linked cluster diagrams or sub--diagrams built with scalar passive bonds only,  with the only exception of the so called {\sl elementary} diagrams, can 
be summed up in closed form by solving the FHNC equations \cite{fantoni_rosati,fabrocini_fantoni}. The contributions associated with the elementary diagrams can be formally included in the FHNC equations, but there is no exact procedure to sum all of them. They can only be taken into account approximatively, by 
explicit calculation of the n--point ($n\geq 4$) basic structures (FHNC/n approximation). However, it is well known that in nuclear matter calculations 
 the FHNC approximation provides very accurate results.
   
On the other hand, diagrams having one or more passive operatorial bonds are calculated at leading order only. Such an approximation is justified by the observation that operatorial correlations are much weaker than the scalar ones. Based on this feature, one would be tempted to conclude that the leading order amounts to  dressing the interaction line with all possible FHNC two--body distribution functions. This is not true as, besides the short range behavior, the 
intermediate range behavior of NN correlations also plays an important role that needs to be taken into account. In particular, tensor correlations, and to some extent also exchange correlations, have a much longer range than the central ones. 

In order to handle this problem, summing the class of chain diagrams turns out to be to be of great importance. To see this, let us consider an extreme example 
of a long range correlation, namely a function $h(r)$ that heals to zero as $\alpha/r^2$, implying that its Fourier transform $\tilde{h}(k)$ behaves as $\beta/k$ in the long wavelength limit. Chain diagrams of $h$--bonds are calculated by means of the convolution integral of the various $h(r_{ij})$ in the chain. In Fourier space convolution integrals are given by products of $\tilde{h}$. One can easily verify that,  in the long wavelength limit, any chain diagram is more singular 
 than $\tilde{h}$. On the contrary, the sum of all the chain diagrams has exactly the same degree of singularity. Hence, summing up the series of chain diagrams takes care of  long range correlations \footnote{This feature is critical to the calculation of the long wavelength limit of the static structure function and the phonon excitations.}.

The above issue is taken care of by summing up the Single Operator Chains (SOC) in the corresponding FHNC/SOC approximation \cite{wiringa_pandha_1,wiringa_pandha_2}. SOC are chain diagrams in which any single passive bond of the chain has a single operator of the type 
$f^c(r_{ij})f^p(r_{ij})\hat{O}^p_{ij}$ or $-h(r_{ij})\ell(k_Fr_{ij})\times P_{ij}$, with $p\leq 6$, or FHNC--dressed versions of them. 
Note that if a single bond of the chain is of the scalar type then the spin trace of the corresponding cluster term vanishes, as the Pauli matrices are traceless.
Then the SOC is the leading order, and at the same time it includes the main features of the long range behavior of tensor and exchange correlations.

The calculation of SOC, as that of FHNC chains, is based upon the convolution integral of the functions corresponding to two consecutive bonds. 
Unlike  FHNC chains, however,  the SOC have operatorial bonds. Therefore,  the basic algorithm is the convolution of two operatorial correlations 
having one common point.  Let us consider two such correlation operators, say $\hat{X}_{ik}=\sum_{p=1,6}x^p(r_{ik})O^{\, p}_{ik}$ and $\hat{Y}_{kj}=\sum_{p=1,6}y^p(r_{kj})O^{\, p}_{kj}$. Their convolution gives rise to a correlation operator of the same algebraic form $\hat{Z}_{ij}=\sum_{p=1,6}z^p(r_{ij})O^{\, p}_{ij}$:
\begin{eqnarray}
\hat{Z}_{ij} &=& \rho\int dx_k \hat{X}_{ik}\hat{Y}_{kj}\, , \nonumber \\
z^r(t_{ij}) &=& \rho\int d\vec{r}_k \xi_{ijk}^{pqr} x^p(r_{ik}) y^q(r_{kj})\, ,
\end{eqnarray}
where the functions $\xi^{pqr}_{ijk}$ depend on the internal angles of the triangle $ijk$. The above equation includes also the convolution of the scalar correlations, which is already taken into account by the FHNC chain equations. Hence, $\xi^{11r}_{ijk}=\delta_{1r}$. If one of the bonds is scalar and the second is operatorial the convolution vanishes, i.e. $\xi^{1qr}_{ijk} = \xi^{p1r}_{ijk} = 0$. The explicit expressions of $\xi^{pqr}_{ijk}$ can be found 
in Refs. \cite{wiringa_pandha_1,fabrocini_fantoni}.

The ordering of the operators within an SOC is immaterial, because the commutator $[\hat{O}_{ik},\hat{O}_{kj}]$  is linear in $\vec{\sigma}_k$ and $\vec{\tau}_k$, and Pauli matrices are traceless. The only orderings that matter are those of passive bonds connected to the interacting points $1$ or $2$. The reason is that  the interaction line may have up to four operators. Therefore, $1$ or $2$ may be reached by up to five operators and one has to take into account the different orderings. The underlying spin algebra is lengthy but straightforward, and it is given in Ref. \cite{wiringa_pandha_1}. As an example, consider the cluster 
diagram of Fig. \ref{fig:ex_cluster_diag} and the corresponding cluster term of Eq. (\ref{eq:ex_cluster_diag}). The two orderings  $\{O^r_{23},O^l_{12}\}O^p_{12}O^q_{12}$ and $O^l_{12}O^p_{12}\{O^q_{12},O^r_{23}\}$ give rise to the same trace, which in the case of $l=p=q=r=2\equiv \sigma$ turns out to be $18$. The full expression of $\langle v\rangle_{(Fig. \ref{fig:ex_cluster_diag})}$ can be easily extracted from the energy term $W_c(de)$ displayed in  Eq. (7.7) of Ref. \cite{wiringa_pandha_1}, and written in terms of the matrices $K^{lpq}$ and $L^{lpq}$ and the vector $A^m$, defined as follows
\begin{eqnarray}
\hat{O}^q_{ij} &=& \sum_{l,p} K^{lpq} \hat{O}^l_{ij} \hat{O}^p_{ij} \, , \nonumber \\
L^{lpq} &=& \pm A^q K^{lpq}\, , \nonumber \\
\text{Tr}_{ij}(\hat{O}^m_{ij} \hat{O}^n_{ij}) &=& \delta_{m,n} A^m \, ,
\end{eqnarray} 
where the $+$ sign applies if 
\begin{equation}
\text{Tr}_{ijk}\big(O^p_{ij}[O^q_{ij},O^r_{jk}]O^l_{ik}\big) = 0\, ,
\end{equation}
while the $-$ sign applies if
\begin{equation}
\text{Tr}_{ijk}\big(O^p_{ij}\{O^q_{ij},O^r_{jk}\}O^l_{ik}\big) = 0\, .
\end{equation}

The $K$--matrices are associated with normal orderings, like $O^p_{ij}O^q_{ij}O^r_{jk}O^l_{ik}$, whereas the $L$--matrices apply to alternate orderings, 
like $O^p_{ij}O^r_{jk}O^q_{ij}O^l_{ik}$.

A second important contribution which is included in FHNC/SOC approximation is the leading order of the vertex corrections. They sum up the 
contributions of sets of subdiagrams which are joined to the basic diagrammatic structure in a single point. Therefore, a vertex correction dresses 
the vertex of all the possible reducible subdiagrams joined to it. The FHNC equations for the full summations of these one--point diagrams are 
given in Ref. \cite{fabrocini_fantoni}. In the FHNC/SOC approximation they are taken into account only at the leading order, i.e. including
single operator rings (SOR), which are nothing but loops of SOC. Vertex corrections play an important role for the fulfillment of the sum rules. 

\subsubsection{Two--body and three--body distribution functions}

The expectation value (\ref{eq:two_body_exp}), can be conveniently rewritten in the form
 
\begin{equation}
\frac{\langle \hat{v} \rangle}{A} =\frac{1}{2}\rho\sum_p \int d\vec{r}_1 d\vec{r}_2 v^{\,p}_{12}\,g^{\, p}_{12} \ ,
\label{eq:2_body_exp_with_g2}
\end{equation}
where
\begin{equation}
g_{12}^p=\frac{A(A-1)}{\rho^2}\frac{\text{Tr}_{12}\int dx_3 \ldots  dx_A \Phi_0^*F^\dagger O_{12}^p F \Phi_0}{\int dX  \ \Phi_0^*F^\dagger F \Phi_0}\, ,
\label{eq:g2_def}
\end{equation}
are the operatorial components of the two--body distribution function.

The FHNC diagrams are divided in 4 separate classes, characterized by the type of bonds ending at the vertices associated with particles $1$ and $2$. The different types 
of vertices are denoted ``d'' for direct, i.e. involving no exchange lines, ``e'' for exchange, i.e. the vertex of an exchange loop, and ``c'' for cyclic, i.e. the vertex 
of an exchange line. Using this notation we can write,
\begin{equation}
g^p= g_{dd}^p + g_{de}^p + g_{ed}^p + g_{ee}^p\, .
\end{equation}
 
The two--body distribution functions satisfy the following sum rules
\begin{eqnarray}
&&\rho\int d\vec{r}_{12} (g^c(r_{12}) -1) = -1\, , \nonumber \\
&&\rho\int d\vec{r}_{12} g^\sigma(r_{12}) = -3\, , \nonumber \\
&&\rho\int d\vec{r}_{12} g^{\sigma\tau}(r_{12}) = 9\, .
\label{eq:sum_rules_distro}
\end{eqnarray}
Note that the above relations also hold true for the distribution functions $g^p$ of the Fermi gas model, as well as for those obtained retaining the $f^c$ correlations only. Another sum rule is given by the expectation value of the kinetic energy, which can be written in three equivalent forms, known as  Pandharipande--Bethe (PB), Jackson--Feenberg (JF) and Clark--Westhaus (CW). In an exact calculation they would all give the same results. Numerical differences between them gauge the degree of accuracy of the approximations employed in the calculation.

The generalization of Eq. (\ref{eq:two_body_exp}) to the case of a three-body potential, e.g. the UIX model, reads
\begin{equation}
\langle V \rangle =\frac{A!}{(A-3)!3!}\frac{\langle \Phi^\dagger_0 |  F^\dagger \hat{V}_{123} F | \Phi_0 \rangle}
{\langle  \Phi^\dagger_0 | F^\dagger F | \Phi_0 \rangle}\, .
\label{eq:exp_three}
\end{equation}

As for the case of the two--body distribution functions $g^p_{12}$, it is useful to define three--body distribution functions $g^p_{123}$, reflecting the operatorial 
structure of $\hat{V}_{123}$ given in Eqs. (\ref{eq:op_structure_UIX}) and (\ref{eq:VR}). Let us write
$\hat{V}_{123}$ as a sum of spin-isospin three--body operators multiplied by scalar functions, depending on the relative distances only
 \begin{equation}
\label{expand:v123}
V_{123} \equiv \sum_p V_{123}^p {O}_{123}^p\, \, .
\end{equation}
From Eqs. (\ref{eq:op_structure_UIX}) and (\ref{eq:VR}) it follows that the sum of the above equation involves 19 operators. The expectation value of $\hat{v}_{123}$ can be written as
 \begin{equation}
\frac{\langle V\rangle}{A}=\frac{1}{3!}\ \rho^2 \sum_P \int dr_{12}dr_{13} V_{123}^p\,g_{123}^p \, ,
\label{eq:3_body_exp_with_g3}
\end{equation}
with
\begin{equation}
g_{123}^p=\frac{A!}{(A-3)!}\frac{\text{Tr}_{123}\int dx_4 \ldots dx_A \Phi^\dagger_0F^\dagger {O}_{123}^p 
F \Phi_0}{ \rho^3\int dX \  \Phi^\dagger_0F^\dagger F \Phi_0}\, .
\label{eq:g3_def}
\end{equation}
 
In Ref.\cite{carlson_pandha_wiringa}, the above expectation value has been computed in FHNC/SOC. The cluster expansion and the corresponding diagrammatic rules of the cluster diagrams are very similar to those outlined  in the case of the two-body potential, with the only difference of three external points, instead of two, all linked by interaction lines.

\begin{figure}[!h]
\vspace{0.2cm}
\begin{center}
\includegraphics[angle=0,width=8.8cm]{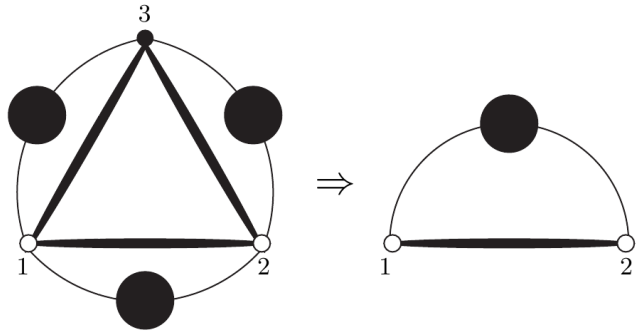}
\caption{Diagrammatic representation of Eq. (\ref{eq:ddp_request}): the two-body density-dependent potential includes the effects of both the bare three-body potential and the correlation and exchange lines. While $g_2$ dresses the line joining particles $1$ and $2$, the dressing being depicted by a line with a big bubble in the middle, $g_3$ dresses the lines $1-2$, $1-3$, and $2-3$. }
\label{fig:g3_g2}
\end{center}
\end{figure}

\section{Derivation of the effective potential}
\label{ddpot}

Our work is aimed at obtaining a two-body density-dependent potential $v_{12}^\rho$ that mimics the three-body 
potential. Hence, our starting point is the requirement that the expectation values of $V_{123}$ and of $v_{12}(\rho)$
be the same:
\begin{equation}
\frac{\langle V \rangle}{A}=\frac{\langle v^{\,\rho}\rangle}{A}\, ,
\end{equation}
implying in turn (compare to  Eqs.(\ref{eq:2_body_exp_with_g2}) and (\ref{eq:3_body_exp_with_g3})) 
\begin{equation}
\sum_{P}\frac{\rho}{3}\int d\vec{r}_3 V_{123}^P\,g_{123}^P=\sum_p v_{12}^{p}(\rho)\,g_{12}^p\, .
\label{eq:ddp_request}
\end{equation}
A diagrammatic representation of the above equation, 
which should be regarded as the definition of  the $v_{12}(\rho)$, 
is shown in Fig. \ref{fig:g3_g2}. The graph on the left-hand side represents the three-body potential times the three-body correlation function, integrated over the coordinates of particle $3$. Correlation and exchange lines are schematically depicted with a line having a bubble in the middle, while the thick solid lines represent the three-body potential. The diagram in the right-hand side 
represents the density-dependent two-body potential, dressed with the two-body distribution function. 
Obviously, $v_{12}^\rho$ has to include not only the three-body potential, but also the effects of correlation and 
exchange lines.  

The left-hand side of Eq.(\ref{eq:ddp_request}) has been evaluated in \cite{carlson_pandha_wiringa} within the FHNC/SOC  scheme. Here we discuss the derivation of the explicit expression of the two-body density-dependent potential appearing in the right-hand side of the equation. The procedure consists of three different step, each corresponding to a different dressing of the diagrams involved in the calculation 

For each of these steps the final result is a density-dependent two-body potential of the form 
\begin{equation}
v_{12}(\rho)=\sum_{p} v^p(\rho,r_{12})O^{p}_{12}\, ,
\label{eq:ddp_potential}
\end{equation}
where, depending on the step, the $v^p(\rho,r_{12})\equiv v_{12}^p(\rho)$ can be expressed in terms of the functions appearing in the definition of the UIX potential, the correlation functions and of the Slater functions.

\subsection*{Step I. Bare approximation}
As a first step in the derivation of the density-dependent potential one integrates the three-body potential over the coordinate of the third particle
\begin{equation}
v_{12}^{\,(I)}(\rho)=\frac{\rho}{3}\int d{x}_3 V_{123}\,.
\label{eq:bare_ddp}
\end{equation}
Diagrammatically the above equation implies that neither interaction nor exchange lines linking particle $3$ with particles $1$ and $2$ are included. Only the two-body distribution function is taken into account in the calculation of the expectation value of $V_{123}$
\begin{equation}
\frac{\langle V \rangle}{A}=\frac{\rho^2}{3!}\sum_p\int d\vec{r}_{12}\Big(\sum_P \int d{x}_3 V_{123}^P\Big)^p g_{12}^p\,.
\end{equation}
Note that only the scalar repulsive term and one permutation of the anticommutator term of the three-body potential provide non vanishing contributions, once the 
trace in the spin--isospin space of the third particle is performed. 

As shown in Fig \ref{fig:compare_potentials}, the contribution of the density-dependent potential to the energy per particle of SNM and PNM $\langle v_{12}^{\,(I)\,\rho}\rangle/A$ is more repulsive than the one obtained from the genuine three-body potential UIX. Thus, the scalar repulsive term is dominant when the three-body potential is integrated over particle $3$.

\subsection{Step II. Inclusion of statistical correlations}

As a second step we have considered the exchange lines that are present both in $g_{123}$ and $g_{12}$. Their treatment  is somewhat complex, and needs to be analyzed in detail. 

\begin{figure}[!h]
\vspace{0.2cm}
\begin{center}
\includegraphics[angle=0,width=4cm]{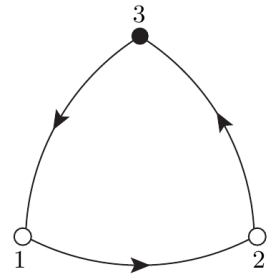}
\caption{Three particle exchange loop.}
\label{fig:3_particle_exchange}
\end{center}
\end{figure}

Consider, for example, the diagram associated with the exchange loop involving particles $1$, $2$ and $3$, 
depicted in Fig. \ref{fig:3_particle_exchange}. 
Its inclusion in the calculation of the density-dependent two-body potential would lead to double counting of exchange lines connecting particles 1 and 2, due to the presence of the  exchange operator $P_{12}$ in $g_{12}$. 
This problem can be circumvented by noting that the antisymmetrization operator acting on particles $1$, $2$ and $3$ can be written in the form
\begin{align}
&1-P_{12}-P_{13}-P_{23}+P_{12}P_{13}+P_{13}P_{23}=\nonumber\\
&\qquad(1-P_{13}-P_{23})\times(1-P_{12})\, ,
\label{eq:exchange_fantoni}
\end{align}
in which the exchange operators contributing to the density-dependent potential only appear in the first term of the 
right-hand side. 

On the other hand, the second term in the right-hand side of Eq. (\ref{eq:exchange_fantoni}) only involves the exchange 
operators $P_{12}$, whose contribution is included in $g_{12}$ and must not be taken into account in the calculation 
of $v_{12}(\rho)$. 

Two features of the above procedure need to be clarified. First, it has to be pointed out that it is exact only within the SOC approximation that allows one to avoid the calculation of commutators between the exchange operators $P_{13}$ and $P_{23}$ and the correlation operators acting on particles $1$ and $2$. The second issue is related to the 
treatment of the radial part of the exchange operators. Although it is certainly true that one can isolate the trace over the spin-isospin degrees of freedom of  particle 3, arising from $P_{13}$ and $P_{23}$, extracting the Slater functions from these 
operators is only possible in the absence of functions depending on the position of particle 3 \cite{pandha_scambi}.

However, this restriction does not apply to the case under consideration, as both the potential and the correlations depend on $\vec{r}_{13}$ and $\vec{r}_{23}$. As a consequence, retaining only the $P_{13}$ and $P_{23}$ exchange operators involves an approximation in the treatment of the the Slater functions, whose validity has been tested by carrying out a numerical calculation.

By singling out  the radial dependence of the exchange operators, and by computing the inverse of the operator $(1-\tilde{P}_{12})$, where $\tilde{P}_{ij}$ denotes the spin-isospin part of $P_{ij}$, it is possible to find a ``Slater Exact'' density-dependent potential $v_{12}^{S.E.}(\rho)$ whose calculation does not involve any approximations concerning the Slater functions. It can be rewritten in the form
\begin{align}
v_{12}^{S.E.}(\rho) &=\frac{\rho}{3}\int d{x}_3 V_{123}\Big\{1+\frac{1}{1-l_{12}^4}\Big[\tilde{P}_{13}(l_{12}^3l_{13}l_{23}-l_{13}^2)\nonumber\\
&\qquad+\tilde{P}_{23}(l_{12}^3l_{13}l_{23}-l_{23}^2)+\tilde{P}_{12}\tilde{P}_{23}(l_{12}l_{13}l_{23}-\nonumber \\
&\qquad l_{12}^2l_{13}^2)+\tilde{P}_{13}\tilde{P}_{23}(l_{12}l_{13}l_{23}-l_{12}^2l_{23}^2)\Big]\Big\}\, ,
\label{eq:exact_exchanges}
\end{align}

\begin{figure}[!t]
\vspace{0.2cm}
\begin{center}
\includegraphics[angle=270,width=9.0cm]{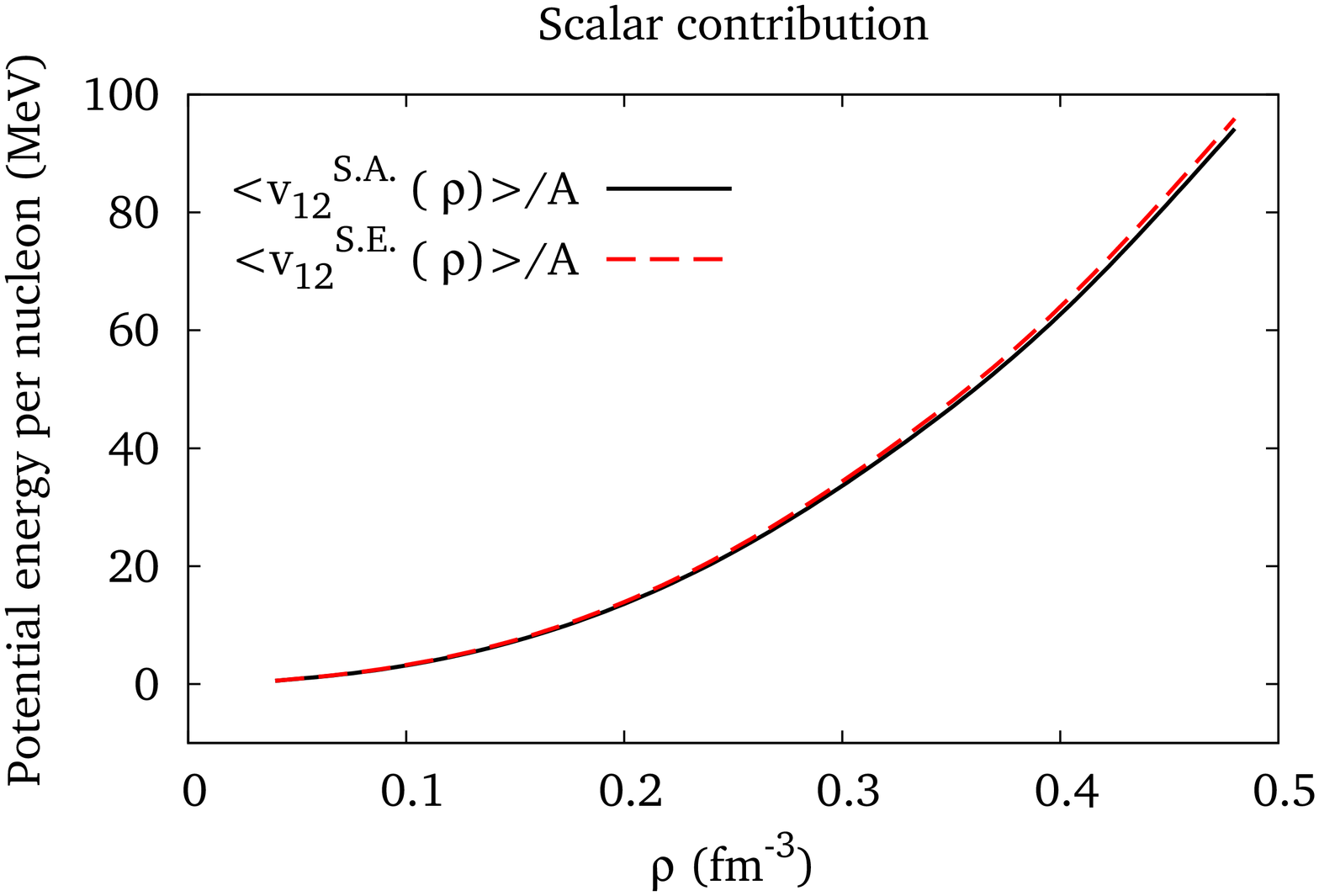}
\vspace{0.1cm}
\includegraphics[angle=270,width=9.0cm]{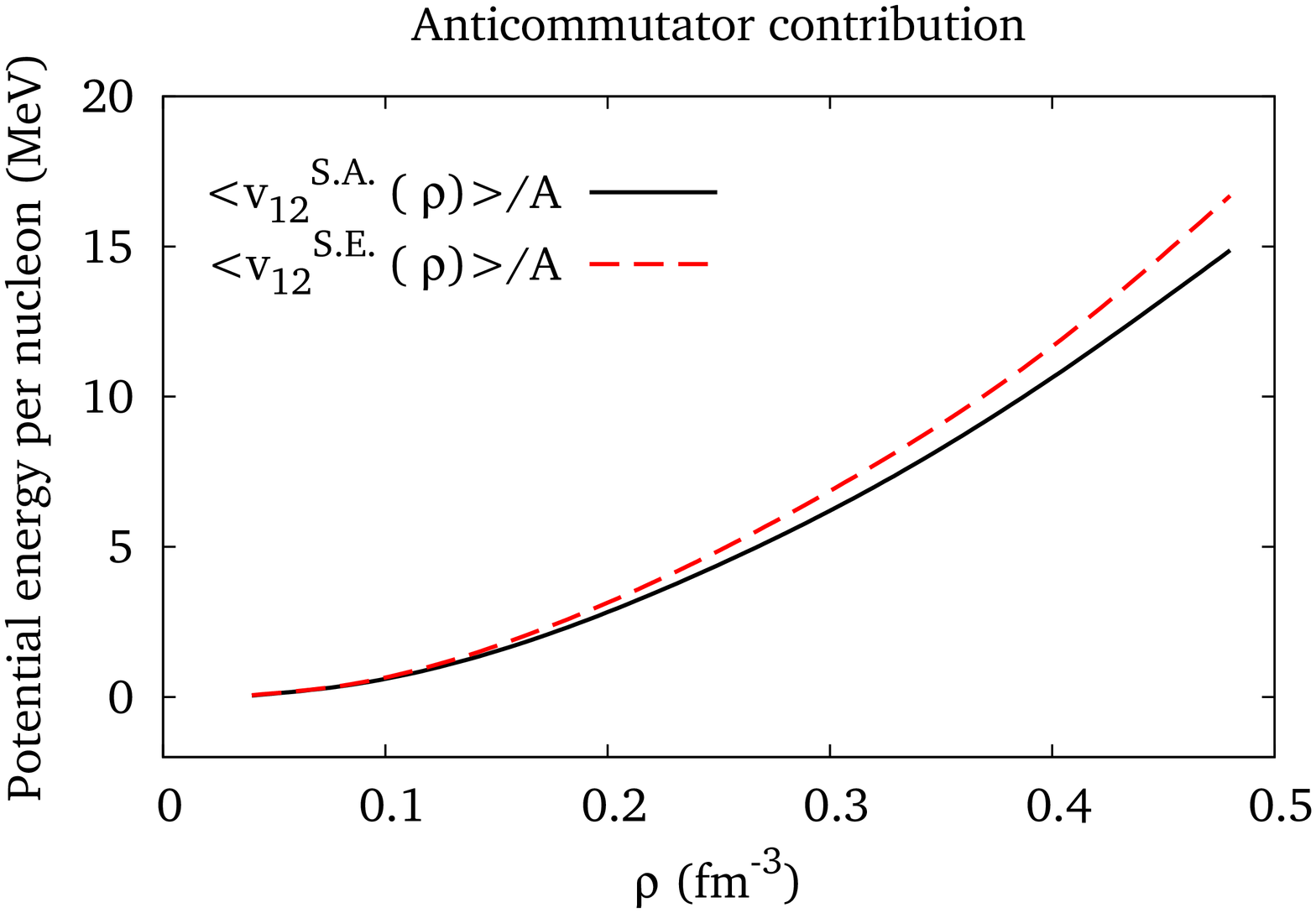}
\caption{Contributions of the density-dependent potential to the energy per particle (see Eqs. (\ref{eq:exact_exchanges}) and (\ref{eq:test_approximate_exchanges})), 
 arising from the scalar term of UIX (upper panel) and from the anticommutator term (lower panel).  \label{fig:exchange_test}}
\end{center}
\end{figure}
Note that in the above equation we have omitted all correlations functions, whose presence is irrelevant to the purpose of our discussion. The density-dependent potential obtained from Eq.({\ref{eq:exact_exchanges}) must be compared to the one resulting from the approximation discussed 
above, which (again neglecting correlations) leads to the expression
\begin{equation}
v_{12}^{S.A.}(\rho)=\frac{\rho}{3}\int d{x}_3 V_{123}(1-\tilde{P}_{13}l_{13}^3-\tilde{P}_{23}l_{23}^2)\, ,
\label{eq:test_approximate_exchanges}
\end{equation}
where ``S. A.'' stands for Slater Approximation.
We have computed $\langle v_{12}^{S.E.}(\rho)\rangle$ and $\langle v_{12}^{S.A.}(\rho)\rangle$ for SNM within the FHNC/SOC scheme, for both the scalar and the anticommutator terms of the UIX potential. 

The results, plotted in Fig. \ref{fig:exchange_test}, clearly show that Eq.(\ref{eq:test_approximate_exchanges}) provides an excellent approximation to the exact result for the exchanges of Eq. (\ref{eq:exact_exchanges}). Hence it has been possible to use Eq. (\ref{eq:test_approximate_exchanges}) also to compute the contribution coming from the commutator of the UIX potential, avoiding the difficulties that would have arisen from an exact calculation of the exchanges. 

The second step in the construction of the density-dependent potential is then 

\begin{equation}
v_{12}^{II }(\rho)\equiv v_{12}^{S.A.}(\rho)\, 
\end{equation}
which is a generalization of the bare potential of Eq. (\ref{eq:bare_ddp}). 

Figure \ref{fig:compare_potentials} shows that taking exchanges into account slightly improves the approximation of the density-dependent potential. However the differences remain large because correlations have not been taken into account.

\subsection{Step III. Inclusion of dynamical correlations}
The third step in the construction of the density-dependent potential amounts to bringing correlations into the game. We have found that the most relevant diagrams are those of Fig. \ref{fig:relevant_diagrams}. 

\begin{figure}[!ht]
\vspace{0.2cm}
\begin{center}
\includegraphics[angle=0,width=5.5cm]{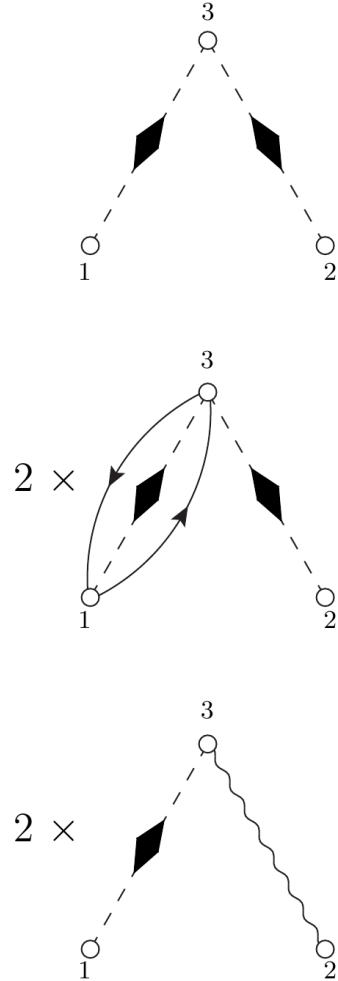}
\caption{Diagrams contributing to the density-dependent potential. The dashed lines with diamonds represent the first order approximation to $g_{bose}^{NLO}(r_{ij})$, discussed in the text. Only diagrams with at most one operator attached to a given point are taken into account. \label{fig:two_body_bose_corr}}
\label{fig:relevant_diagrams}
\end{center}
\end{figure}

Note that, in order to simplify the pictures, all interaction lines are omitted. However, it is understood that the three-body potential  is acting on particles $1$, $2$ and $3$. Correlation and exchange lines involving these particles are depicted as if they were passive interaction lines. Moreover, in order to include higher order cluster terms, we have replaced the scalar correlation line ${f^c_{ij}}^2$ with the Next to Leading Order (NLO) 
approximation to the bosonic two-body correlation function:
\begin{equation}
{f^{c}_{ij}}^2\rightarrow g_{bose}^{NLO}(r_{ij})={f^{c}_{ij}}^2\Big(1+\rho\int d\vec{r}_3 h_{13} h_{23}\Big)\, .
\label{eq:g12_NLO}
\end{equation}
The full bosonic $g_{bose}(r_{ij})$ or $g_{dd}(r_{ij})$ might be used instead of the NLO approximation. However, including higher order terms would have broken our cluster expansion. The correction to ${f^{c}_{ij}}^2$ of Eq. (\ref{eq:g12_NLO}), whose diagrammatic representation is displayed in Fig. \ref{g_2_bose}, can indeed be considered to be of the same order as the operatorial correlations.  

Figure \ref{fig:relevant_diagrams} shows that the vertices corresponding to particles 1 and 2 are
not connected by either correlation or exchange lines. All connections allowed by the diagrammatic rules are 
taken into account multiplying the density-dependent potential by the two-body distribution function, according 
to the definition of Eq.(\ref{eq:ddp_request}).
 
We have already discussed the exchange lines issue, coming to the conclusion that only the exchanges $P_{13}$ and $P_{23}$ have to be taken into account. This is represented by the second diagram, where the factor $2$ is due to the symmetry of the three-body potential, that takes into account both $P_{13}$ and $P_{23}$.  

\begin{figure}[!t]
\vspace{0.2cm}
\begin{center}
\includegraphics[angle=0,width=9cm]{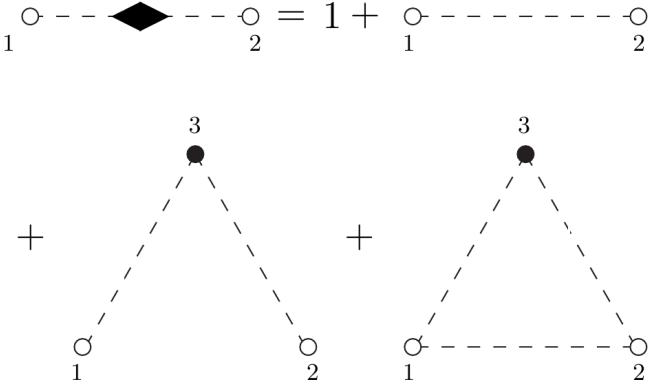}
\caption{NLO approximation to the bosonic two-body correlation function.}
\label{g_2_bose}\end{center}
\end{figure}

The explicit expression of $v_{12}^{(III)}(\rho)$ obtained including the diagrams of 
Fig. \ref{fig:relevant_diagrams} can be cast in the form
\begin{align}
v^{(III)}_{12}(\rho)&=\frac{\rho}{3}\int d x_3\,V_{123}\,\Big[g_{bose}^{NLO}(r_{13}) g_{bose}^{NLO}(r_{23})\nonumber \\
&\times(1-2P_{13}\ell_{13}^2)+4g_{bose}^{NLO}(r_{13})f_{c}(r_{23})\hat{f}(r_{23})\Big]\, ,
\label{eq:ddp_diagrams_expression}
\end{align}
where $\hat{f}(r_{23})$ denotes the sum of non central correlations
\begin{equation}
\hat{f}(r_{23})=\sum_{p\neq 1}^6 f^p(r_{23})O^{p}_{ij}\, .
\end{equation}
Note that, in principle,  an additional term involving the anticommutator between the potential and the 
 correlation function should appear in the second line of the above equation. However, due to the structure 
 of the potential it turns out that 
\begin{equation}
\int d{x}_3 \{V_{123},\hat{f}(r_{23})\}=2\int d x_{3} V_{123} \hat{f}(r_{23})\, .
\end{equation}

The calculation of the right-hand side of  of Eq. (\ref{eq:ddp_diagrams_expression}) requires the evaluation of the traces of commutators and anticommutators of spin-isospin operators, as well as the use of suitable angular functions needed to 
carry out the integration over $\vec{r}_3$. 

\begin{figure}[!h]
\begin{center}
\includegraphics[angle=270,width=9.0cm]{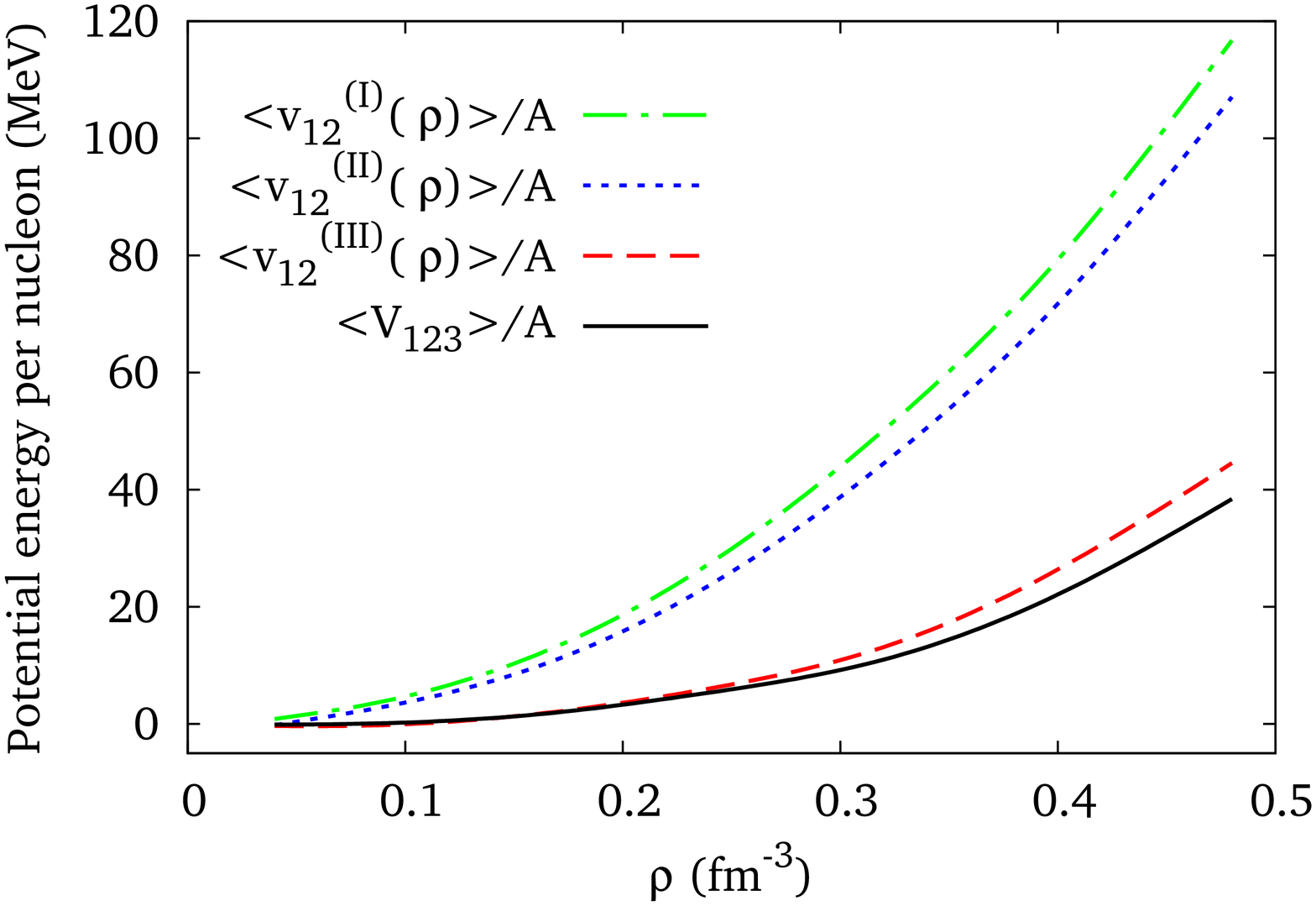}
\vspace{0.1cm}
\includegraphics[angle=270,width=9.0cm]{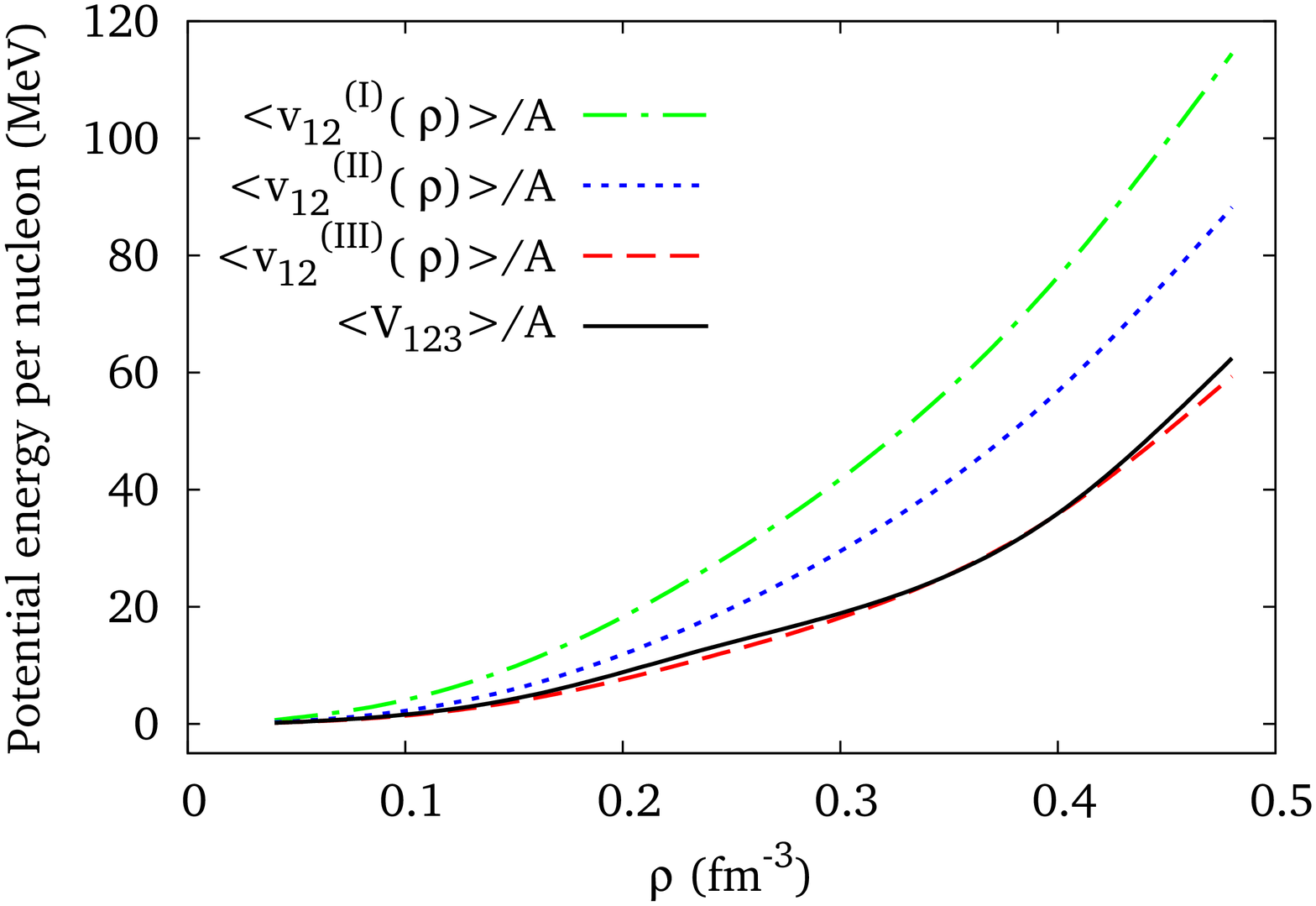}
\vspace{0.1cm}
\caption{Contributions of the density-dependent potential to the energy per particle of SNM (upper panel) and PNM (lower panel), compared to 
the expectation value of the genuine three-body potential UIX: $\langle V_{123}\rangle/A$. \label{fig:compare_potentials}}
\end{center}
\end{figure}

As for the previous steps, we have computed the contribution of the density-dependent potential $v_{12}^{(III)}(\rho)$ to the energy per particle. The results 
of Fig. \ref{fig:compare_potentials} demonstrate that the density-dependent potential including correlations is able to reproduce the results 
obtained using genuine three-body UIX to remarkable accuracy. 

To simplify the notation, at this point it is convenient to identify
\begin{equation}
 v_{12}(\rho) \equiv v_{12}^{(III)}(\rho)\, .
\end{equation}

Note that the above potential exhibits important differences when acting in PNM and in SNM. For example, in SNM $v^p(\rho,r_{12}) \neq 0$ for $p=1,\sigma_{12}\tau_{12},S_{12}\tau_{12}$, while in PNM $v^p(\rho,r_{12}) \neq 0$ for $p=1,\sigma_{12},S_{12}$.

\section{Numerical Calculations}
\label{Num_Calc}
\subsection{Variational approach in FHNC/SOC approximation}
\label{Variational_SA}
An upperbound  to the binding energy per particle, $E_V/A$, can be obtained by using the variational method, which amounts to minimizing the energy expectation value $\langle H\rangle/A$ with respect to the variational parameters included in the model.  Its cluster expansion is given by
\begin{equation}
\frac{\langle H\rangle}{A} = T_F +  (\Delta E)_2 + \makebox{higher order terms} \ ,
\label{eq:energy_cluster_cont}
\end{equation}
 where $T_F$ is the energy of the non interacting Fermi gas and $(\Delta E)_2$ denotes the contribution of two-nucleon 
clusters, described by the diagram of Fig \ref{fig:2_body_cc}. Neglecting higher order cluster contributions, the functional minimization of $\langle H\rangle/A$ leads to a set of  six Euler-Lagrange equations, to be solved with proper constraints that force $f^c$ and $f^{(p>1)}$ to ``heal'' at one and zero, respectively. That is most efficiently achieved 
through the boundary conditions \cite{LP1,wiringa_pandha_1}
\begin{eqnarray}
f^{p}(r\geq d^p) &=& \delta_{p1}\, , \nonumber \\
\frac{df^p(r)}{dr}\mid_{d^p} &=& 0\, . 
\end{eqnarray}
Numerical calculations are generally carried out using only  two independent ``healing distances'': $d_c~=~d^{p=1 \dots 4}$ and $d_t=d^{5,6}$.

\begin{figure}[!h]
\vspace{0.2cm}
\begin{center}
\includegraphics[angle=0,width=9cm]{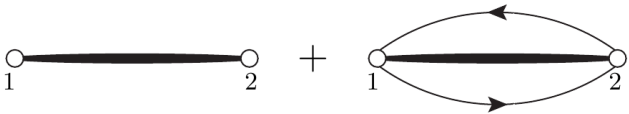}
\caption{Diagrammatic representation of the two-body cluster contribution $(\Delta E)_2$ of Eq. (\ref{eq:energy_cluster_cont}). The thick lines represents both the potential and a kinetic contribution, involving derivatives  acting only on the correlation functions. The effect of the other derivatives is included in $T_F$.}
\label{fig:2_body_cc}
\end{center}
\end{figure}

Additional and important variational parameters are the quenching factors $\alpha_p$ whose introduction simulates modifications of the two--body potentials entering in the Euler--Lagrange differential equations arising from the screening induced by the presence of the nuclear medium 
\begin{equation}
v_{ij}=\sum_{p=1}^6 \alpha_p v^{p}(r_{ij})O^{p}_{ij}\, .
\end{equation}
The full potential is, of course, used in the energy expectation value.
In addition,  the resulting correlation functions $f^p$ are often rescaled according to  
\begin{equation}
F_{ij}=\sum_{p=1}^6 \beta_p f^{p}(r_{ij})O^{p}_{ij}\; ,
\end{equation}
 
The energy expectation value $\langle H\rangle/A$, calculated in full FHNC/SOC approximation is minimized with respect to variations of $d_c$, $d_t$, $\beta_{p}$, and $\alpha_{p}$.

To determine the best values of the variational parameters we have used a version of the ``Simulated annealing'' algorithm \cite{SA}. 
In metallurgy the annealing procedure consists in heating and then slowly cooling a metal, to decrease the defects of its structure. During the heating the atoms gain kinetic energy and move away from their initial equilibrium positions, passing through states of higher energy. Afterwards, when the metal slowly cools, it is possible that the atoms freeze in a different configuration with respect to the initial one, corresponding to a lower value of the energy.

In minimization problems the analog of the position of the atoms are the values of the parameters to be optimized, in our case $d_c$, $d_t$, $\beta_{p}$ and $\alpha_p$, while the energy of the system correspond to the function that has to be minimized, that in our case is the variational energy
\begin{equation}
E_V=E_V(d_c,d_t,\beta_{p},\alpha_p)\, .
\end{equation}

In the simulated annealing procedure, the parameters $d_c$, $d_t$, $\beta_{p}, \alpha_p$ are drawn from the Boltzmann distribution, $\exp(-E_{V}/T)$, where $T$ is just a parameter of the simulated annealing algorithm, having no physical meaning.

We have used a Metropolis algorithm, with acceptance probability of passing from the state $s=\{d_c,d_t,\beta_{p},\alpha_p\}$ to the proposed state $s'=\{d_c',d_t',\beta_{p}',\alpha_p'\}$ given by
\begin{equation}
P_{s,s'}=\exp\Big[-\frac{E(s')-E(s)}{T}\Big]\, ,
\end{equation}
By looking at the distribution of the parameters resulting from the Metropolis random walk, it is possible to find the values $\tilde{d}_c$, $\tilde{d}_t$, $\tilde{\beta}_{p}$ and $\tilde{\alpha}_p$ corresponding to the minimum of $E_V$, e.g. to the maximum of the Boltzmann distribution. As the fictitious temperature $T$ is lowered, the system approaches the equilibrium and the values of the parameters get closer and closer to $\tilde{d}_c$, $\tilde{d}_t$, $\tilde{\beta}_{p},\tilde{\alpha}_p$ .

\begin{figure}[!h]
\vspace{0.2cm}
\begin{center}
\includegraphics[angle=270,width=9.0cm]{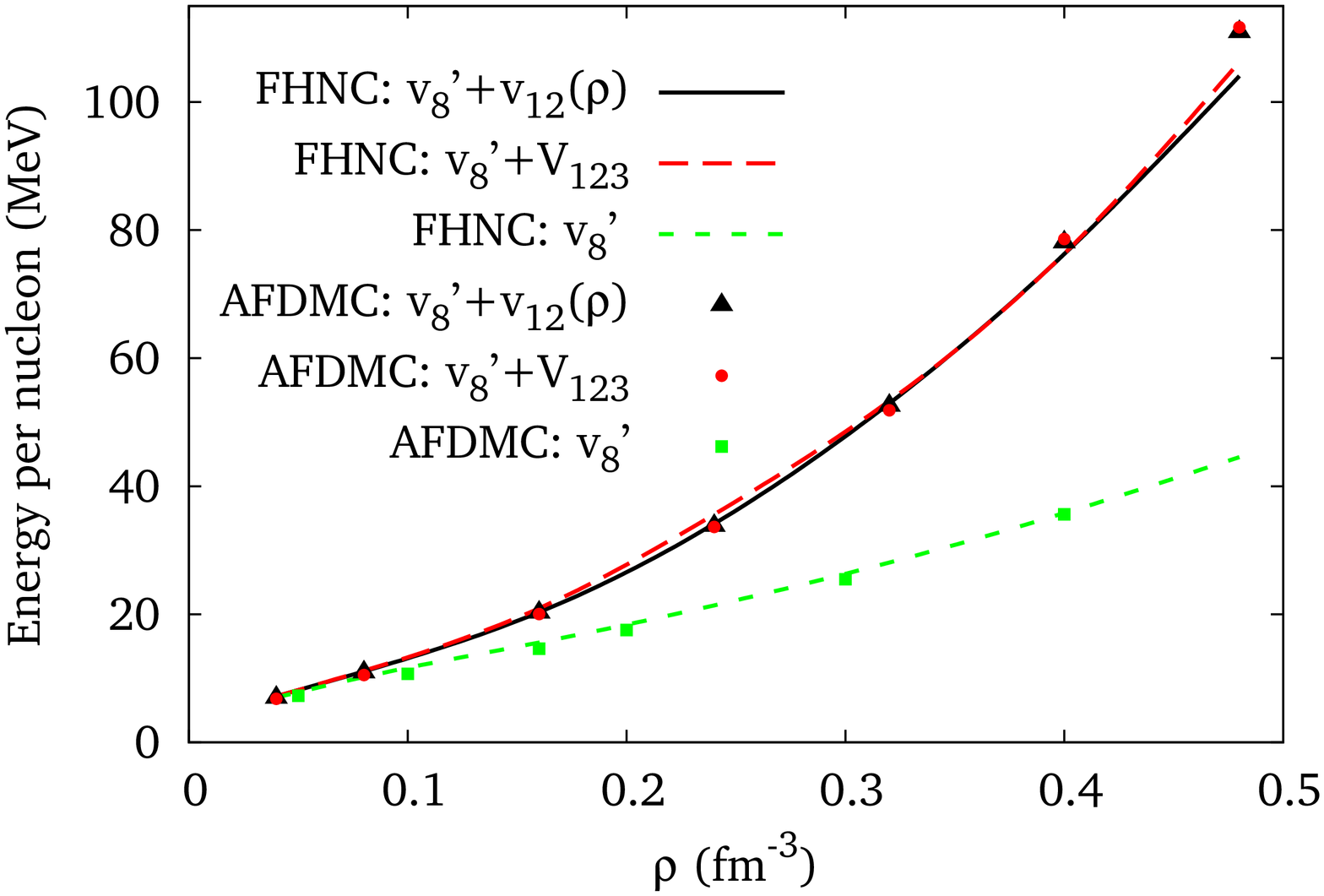}
\vspace{0.1cm}
\includegraphics[angle=270,width=9.0cm]{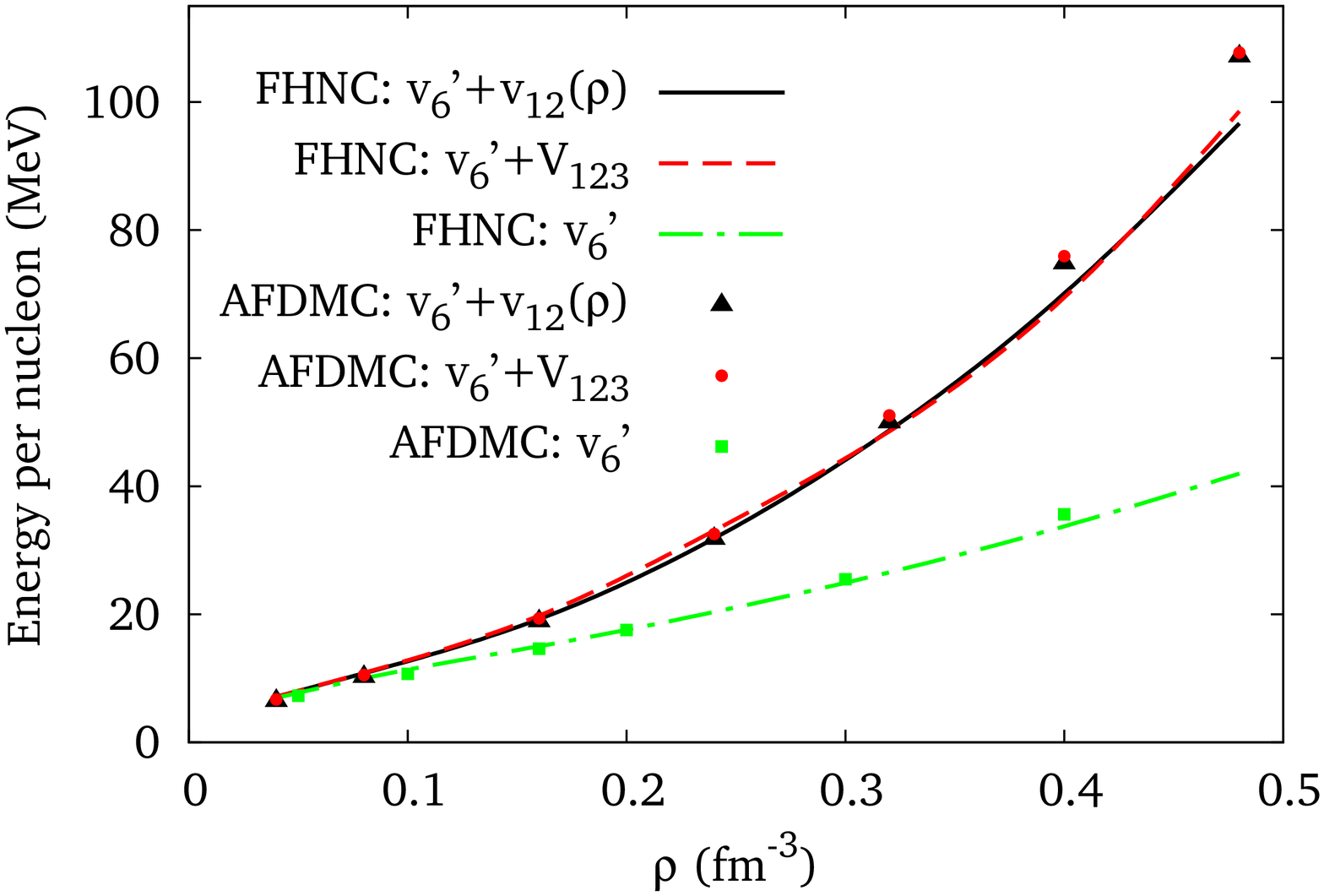}
\vspace{0.1cm}
\caption{Energy per particle for PNM, obtained using the density-dependent potential of Eq. (\ref{eq:bare_ddp}) added to the Argonne $v_{8}^\prime$ (upper panel) 
and to Argonne $v_{6}^\prime$ (lower panel) potentials. The energies are compared to those obtained from the genuine three-body potential and from 
the two-body potentials alone.  \label{fig:eos_pnm}}
\end{center}
\end{figure}

The best solution found during the execution of the Metropolis algorithm has been kept.
The discrete values of the temperature, $T_i$, as well as the numbers of Monte Carlo steps for each $T_i$ have been chosen in such a way that different executions of the simulated annealing procedure produce the same value for $\tilde{d}_c$, $\tilde{d}_t$, $\tilde{\beta}_{p}$ and $\tilde{\alpha}_p$.

A constrained optimization has been performed, by imposing the sum rules for the kinetic energy and for the scalar two-body distribution function. In particular the difference between the Pandharipande-Bethe (PB) and the Jackson-Feenberg (JF) kinetic energies has been forced to be less than $10\, \%$ of the Fermi Energy $T_F$ of Eq. (\ref{eq:energy_cluster_cont}), while the sum rule (\ref{eq:sum_rules_distro}) for $g^c(r_{12})$ has been satisfied with a precision of $3\, \%$.

In our calculations we have optimized the variational paremeters for four different Hamiltonians, each corresponding to different potential terms: Argonne $v_{8}^\prime$, Argonne $v_{8}^\prime\,+\, $UIX, Argonne $v_{6}^\prime$, and Argonne $v_{6}^\prime\,+\, $UIX.

The energy per particle of SNM and PNM computed adding to the two-body potentials Argonne $v_{8}^\prime$ and Argonne $v_{6}^\prime$ the density-dependent potential of Eq. (\ref{eq:ddp_potential}), have been compared to the results obtained using the hamiltonian of Eq.(\ref{hamiltonian}) with the same two-body potentials and the Urbana IX three-body potential. The energy associated with the density-dependent potential has been computed with the same variational parameters resulting from the genuine three-body potential, i. e. no optimization procedure has been performed for the density-dependent potentials. 

Both calculations have been consistently carried out within the FHNC/SOC scheme.

It is worth noting that our simulated annealing constrained optimization allows us to: i) reduce the violation of the variational principle due to the FHNC/SOC approximation; ii) perform an accurate scan of the parameter space. As a consequence, our FHNC/SOC calculations provide very close results to those obtained via Monte Carlo calculations, as shown in Figs. \ref{fig:eos_pnm} and \ref{fig:eos_snm}, to be compared with those of Ref. \cite{AFDMC1} where the agreement between FHNC and Monte Carlo methods were not nearly as good.

\subsection{Auxiliary Field Diffusion Monte Carlo (AFDMC) approach}

In order to check the validity of our variational FHNC/SOC calculations, we carried out AFDMC simulations \cite{Schmidt1999} for both PNM and SNM.

The AFDMC method has proved to be a powerful approach to deal with large nuclear systems, such as medium--heavy nuclei and infinite matter.
Using a fixed-phase like approximation, AFDMC also yields 
results in very good agreement with those obtained from Green Function Monte Carlo (GFMC) calculations for light nuclei \cite{Gandolfi2007b}.

We have computed the equation of state of PNM and SNM using the AFDMC method with the fixed-phase like approximation. 
We simulated PNM with $A=66$ and SNM with $A=28$ nucleons in a periodic box, as described in \cite{Gandolfi2009} and \cite{Gandolfi2008}.
The finite-size errors in PNM simulations have been investigated in \cite{Gandolfi2008} by comparing the twist averaged boundary conditions with the periodic box 
condition. It is remarkable that the energies of 66 neutrons computed using either twist averaging or periodic boundary conditions turn out to be almost the same. 
This essentially follows from the fact that the kinetic energy of 66 fermions approaches the thermodynamic limit very well. The finite-size corrections due to 
the interaction are correctly estimated by including the contributions given by neighboring cells to the simulation box\cite{Sarsa2003}.
From the above results for PNM and those reported in \cite{AFDMC1} for SNM, we can estimate that the finite size errors in the present AFDMC calculations 
do not exceed 1\% of the energy.

The statistical errors, on the other hand, are very small and in the Figures are always hidden by the squares, the triangles and the circles representing the AFDMC energies.


\subsection{PNM equations of state}
In the PNM case (see Fig. \ref{fig:eos_pnm}), the EoS obtained with the three-body potential UIX and using the density-dependent two-body potential are very close to 
each other. For comparison, in  Fig. \ref{fig:eos_pnm} we also report the results of calculations carried out including the two--body potential only. 
In our approximation, with the exception of the line with diamonds of Fig. \ref{fig:two_body_bose_corr}, we have neglected the cluster contributions proportional to $\rho^2$. One could then have guessed that the curves corresponding to the UIX and density-dependent potential would have slightly moved away from each other at higher densities because, as the density increases, the contributions of higher order diagrams become more important. Probably, in this case a compensation among these second and higher order terms takes place.

The density-dependent potential obtained in the FHNC/SOC framework has been also employed in AFDMC calculations. As can be plainly seen in Fig. \ref{fig:eos_pnm}, the triangles representing the results of  this calculation are very close, when not superimposed, to the circles corresponding to the UIX three-body potential AFDMC results.

\subsection{SNM equation of state}
In the EoS of symmetric nuclear matter, the above compensation does not appear to occur, as can be seen in Fig. \ref{fig:eos_snm}. At densities lower than $\rho=0.32 \ \text{fm}^{-3}$, the curves resulting from UIX and the density-dependent potential are very close to one other, while for  $\rho>0.32 \text{fm}^{-3}$ 
a gap between them begins to develop.

The gap is smaller when the two-body potential Argonne $v_{8}^\prime$ is used, but the reason for this is not completely clear.

\begin{figure}[!h]
\vspace{0.2cm}
\begin{center}
\includegraphics[angle=270,width=9.0cm]{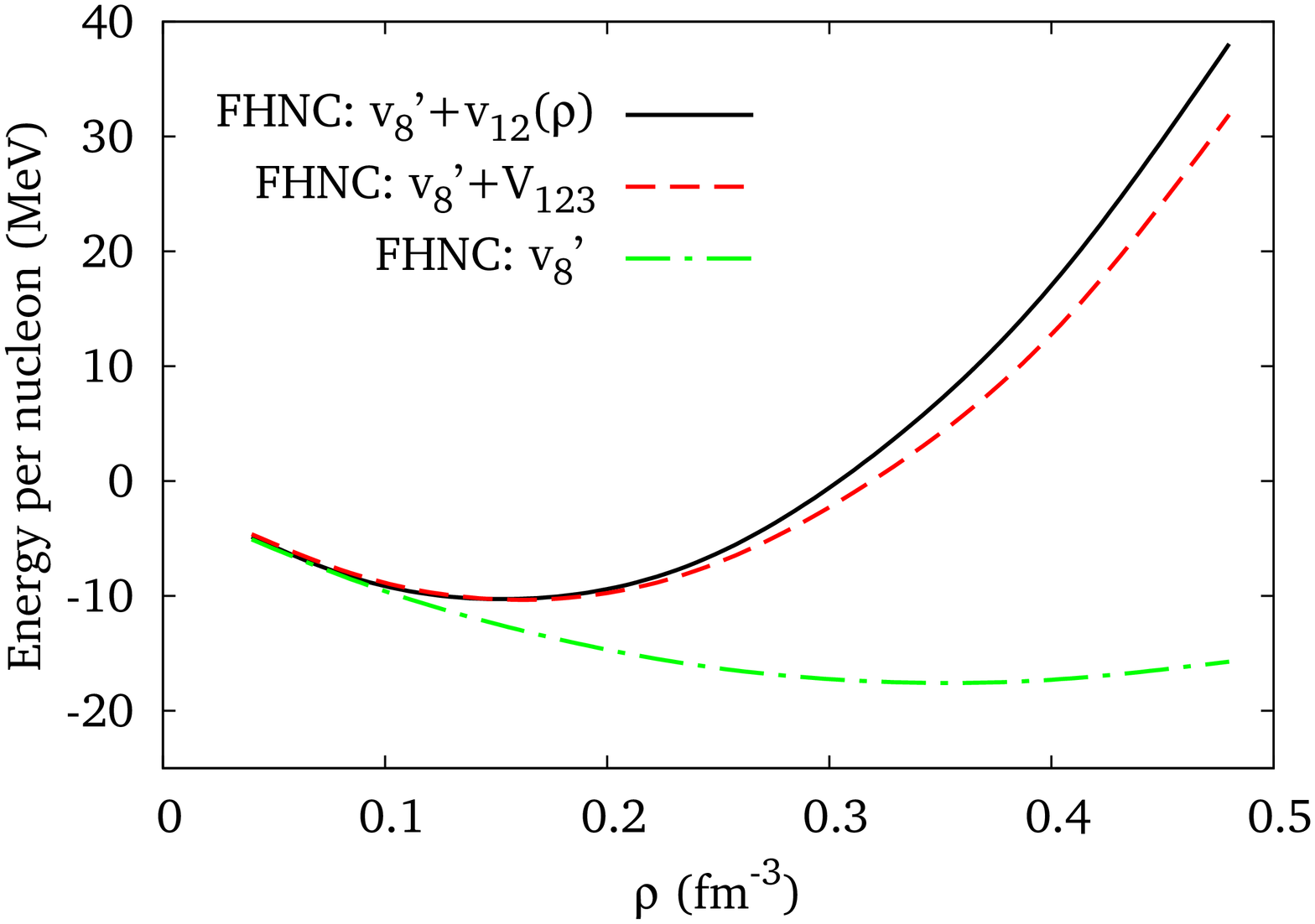}
\vspace{0.1cm}
\includegraphics[angle=270,width=9.0cm]{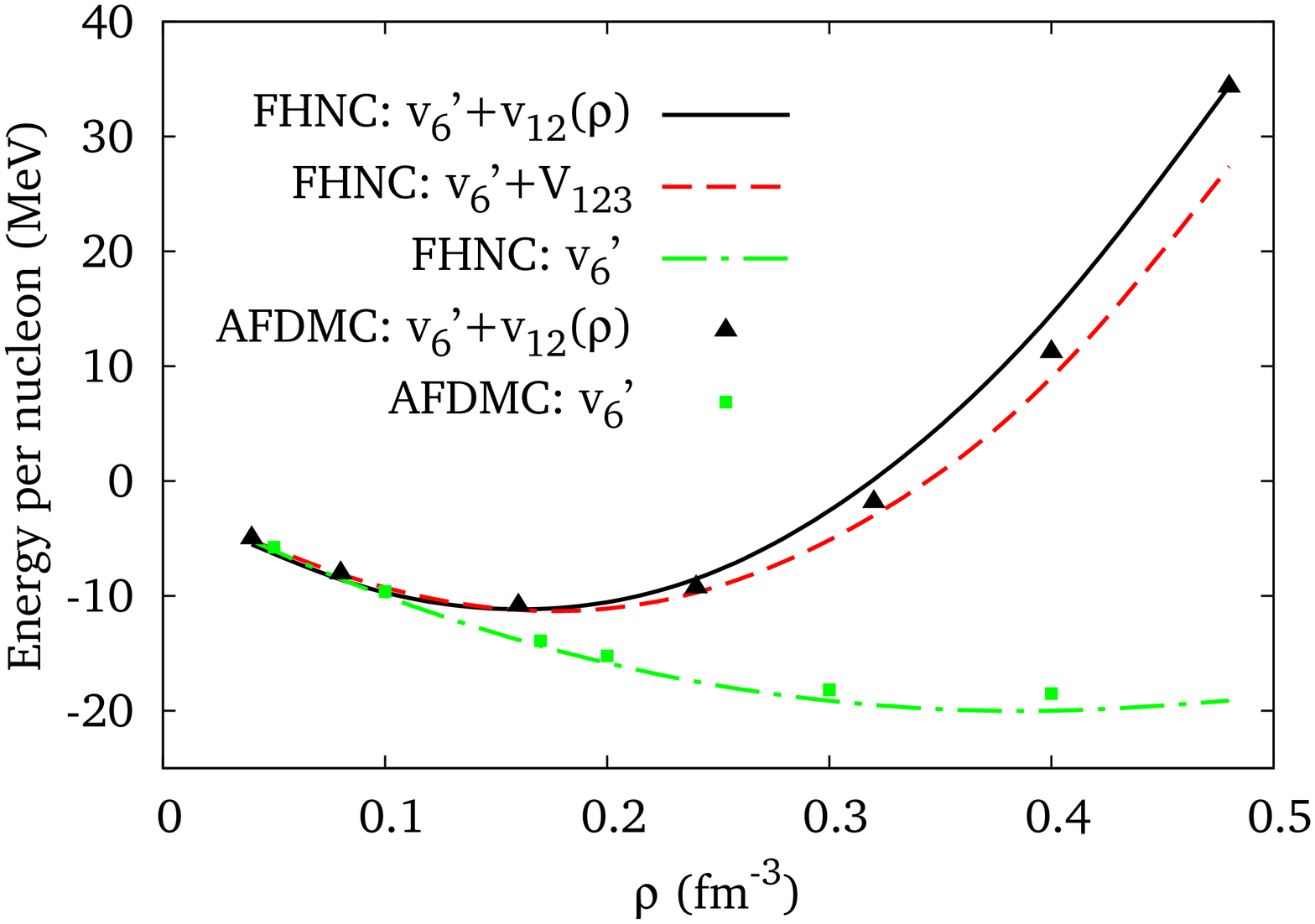}
\vspace{0.1cm}
\caption{Same as in Fig. \ref{fig:eos_pnm}, but for SNM
\label{fig:eos_snm}}
\end{center}
\end{figure}

We have computed the saturation density $\rho_0$, the binding energy per particle $E(\rho_0)$ and the compressibility $K = 9\rho_0 (\partial E(\rho)/\partial\rho)^2$ for all the EoS of Fig. \ref{fig:eos_snm}. The variational FHNC/SOC results are listed in Table \ref{table:parameters_eos}, while those coming from the AFDMC calculation with  $v_{6}^\prime+v_{12}(\rho)$ potential are: $\rho_0=0.17\,\text{fm}^{-3}$, $E_0=-10.9\,\text{MeV}$ and K$=201\,\text{MeV}$.
\begin{table}[]
\caption{Values for the saturation densities, the binding energy per particle, and the compressibility of SNM obtained from the variational FHNC/SOC EoS of Fig. \ref{fig:eos_snm}. \label{table:parameters_eos}}
\vspace{0.3cm}
\begin{tabular}{c c c c c} 
\hline 
 	& $v_{6}^\prime + V_{123}\quad$ & $ v_{6}^\prime + v(\rho) $ & $ v_{8}^\prime + V_{123} $ & $ v_{8}^\prime + v(\rho)$\\ 
\hline
$\rho_0$ (fm$^{-3}$)& 0.17 & 0.16 & 0.16 & 0.15 \\ 

$E_0$ (MeV) &-11.3 & -11.2 & -10.3 & -10.3 \\ 

K (MeV) & 205 & 192 & 189 & 198 \\ 
\hline
 
\end{tabular} 
\vspace{0.1cm}
\end{table}

The saturation densities are quite close to the empirical value $\rho_0=0.16\,\text{fm}^{-3}$ (MeV). For the genuine three-body potential this is not surprising, since the parameter $U_0$ is chosen to fit the saturation density, as discussed in Section \ref{TBF}. On the other hand,  the fact that the density-dependent potential also reproduces 
this value is remarkable and needs to be emphasized.

The binding energies obtained with $v_{12}(\rho)$ are very close to those coming from UIX potential, but they are larger than the empirical value $E_0=-16\,\text{MeV}$. 

As for the compressibility, the experimental value $K\approx 240\,\text{MeV}$ suffers of sizable uncertainties. However, also in this case the result obtained with the density-dependent potential differs from that obtained with the UIX potential by less than $5 \%$.

\section{Conclusions}
\label{conclusions}

We have developed a novel approach, allowing one to obtain an effective density-dependent NN potential 
taking into account the effects of three-- and many-- nucleon interactions. 

The resulting effective potential can be easily used in calculations of nuclear 
properties within many-body approaches based on phenomenological hamiltonians, including the effects of strong NN correlations, which can not be 
treated in {\em standard} perturbation theory in the Fermi gas basis. Moreover, the derivation of the density-dependent NN potential is fully consistent 
with the treatment of correlations underlying the FHNC and AFDMC approaches. 

While the reduction of $n$-body potentials to a two-body density-dependent potential is reminiscent of the approach of Refs. \cite{LP1,FrP}, 
our scheme significantly improves upon the TNI model, in that i) it is based on a microscopic model of the three nucleon interaction providing 
a quantitative description  of the properties of few nucleon systems and ii) allows for a consistent inclusion of higher order terms in the 
density expansion, associated with four- and more-nucleon interactions. 

As shown in Figs. \ref{fig:eos_pnm} and \ref{fig:eos_snm}, the results of calculations of the PNM and SNM equation of state carried out using the density-dependent potential turn out to be very close to 
those obtained with the UIX three-body potential. In this context, a critical role is played by the treatment of both dynamical and statistical correlations, 
whose inclusion brings the expectation value of the effective potential into agreement with that of the UIX potential (see Fig.~\ref{fig:compare_potentials}).
This is a distinctive feature of our approach, as compared to different reduction schemes based on effective interactions,  suitable for use in standard perturbation 
theory \cite{hebeler_schwenk,weise}.

Using the density-dependent potential we have been able to carry out, for the first time, a AFDMC calculation of the equation of state of SNM consistently including the effects of three nucleon forces. The results of this calculation show that the $v_6^\prime + $UIX hamiltonian, or equivalently the one including 
the effective potential, fails to reproduce the empirical data. 

The FHNC results obtained using the $v_8^\prime$ potential indicate that the 5--6~MeV underbinding at equilibrium density can not be accounted for replacing the $v_6^\prime$ with a more refined model, such as $v_{18}$.  Hence, the discrepancy has to be ascribed either to deficiencies of the UIX model or to the effect of interactions involving more than three nucleons. 

The immediate follow up of our work is the AFDMC calculation of the SNM equation of state with the $v_8^\prime$ potential and the density-dependent potential, which is currently being carried out. Further development will include a study of the dependence on the specific model of three-nucleon force, as well as the inclusion of of four- and many-nucleon interactions, whose effects are expected to be critical for the determination of the 
properties of high density neutron star matter.

As a final remark, the effective potential discussed in our paper could be easily employed in many-body approaches other than 
those based on the CBF formalism or quantum  Monte Carlo simulations, such as the G-matrix and self-consistent Green function theories \cite{Baldo, Polls, 
Dickhoff}.
\vspace{1cm}
\begin{acknowledgements}
KES was partially supported by NSF grant PHY-0757703.
\end{acknowledgements}

\end{document}